\documentclass[letterpaper,twocolumn,10pt]{article}
\usepackage{usenix}

\usepackage{amsmath}

\usepackage{graphicx}
\usepackage{float} 
\usepackage{subfigure}
\usepackage{multirow}
\usepackage{booktabs}
\usepackage{siunitx}
\usepackage{xspace}
\usepackage{listings}
\hypersetup{colorlinks=true,citecolor=blue,filecolor=black,linkcolor=blue,urlcolor=black}

\definecolor{codegreen}{rgb}{0,0.6,0}
\definecolor{codegray}{rgb}{0.5,0.5,0.5}

\lstdefinestyle{codestyle}{
    commentstyle=\color{codegreen},
    keywordstyle=\color{black}\bfseries,
    numberstyle=\color{codegray},
    stringstyle=\color{codepurple},
    basicstyle=\ttfamily\mdseries\scriptsize,
    emph={MeshInsight,profiler},
    emphstyle={\color{magenta}\bfseries},
    breakatwhitespace=false, 
    frame=bottomline,        
    breaklines=true,                 
    captionpos=b,                    
    keepspaces=true,                 
    numbers=left,                    
    numbersep=5pt,                  
    showspaces=false,                
    showstringspaces=false,
    showtabs=false,                  
    tabsize=2
}

\def\Snospace~{\S{}}

\newcommand{\sys}{MeshInsight\xspace}

\newcommand{\autorefsuffix}[2]{\hyperref[#1]{\autoref*{#1}#2}}

\ifdefined\isFinalized
  \newcommand{\note}[1]{}
  \newcommand{\todo}[1]{}
  \newcommand{\Danyang}[1]{}
  \newcommand{\Guozhen}[1]{}
  \newcommand{\xz}[1]{}
  \newcommand{\matt}[1]{}
  \newcommand{\ratul}[1]{}
  \newcommand{\arvind}[1]{}
\else
  \newcommand{\note}[1]{{\color{red}[Note: #1]}}
  \newcommand{\todo}[1]{{\color{red}[TODO: #1]}}
  \newcommand{\Danyang}[1]{{\color{red}[Danyang: #1]}}
  \newcommand{\Guozhen}[1]{{\color{blue}[Guozhen: #1]}}
  
  \newcommand{\matt}[1]{{\color{teal}[Matt: #1]}}
  \newcommand{\ratul}[1]{{\color{orange}[Ratul: #1]}}
  \newcommand{\arvind}[1]{{\color{olive}[Arvind: #1]}}
\fi

\newcommand{\OMIT}[1]{}

\newcommand{\paraspace}{\vspace{0.05in}}
\newcommand{\parab}[1]{\paraspace\noindent{\bf #1} }

\begin{document}
\date{}

\title{\Large \bf Dissecting Service Mesh Overheads}

\author{
\rm{Xiangfeng Zhu$^{\text{1}}$ \enskip
    Guozhen She$^{\text{2}}$ \enskip
    Bowen Xue$^{\text{1}}$ \enskip}\\
\rm{Yu Zhang$^{\text{3}}$ \enskip
    Yongsu Zhang$^{\text{3}}$ \enskip
    Xuan Kelvin Zou$^{\text{3}}$ \enskip
    Xiongchun Duan$^{\text{3}}$ \enskip
    Peng He$^{\text{3}}$ \enskip}\\
\rm{Arvind Krishnamurthy$^{\text{1}}$ \enskip
    Matthew Lentz$^{\text{2}}$ \enskip
    Danyang Zhuo$^{\text{2}}$ \enskip
    Ratul Mahajan$^{\text{1,4}}$ \enskip}\\
\\
  {$^{\text{1}}$University of Washington\enskip $^{\text{2}}$Duke University\enskip $^{\text{3}}$Bytedance\enskip $^{\text{4}}$Intentionet} \\
}

\maketitle
\section*{Abstract}
Service meshes play a central role in the modern application ecosystem by providing an easy and flexible way to connect different services that form a distributed application. However, because of the way they interpose on application traffic, they can substantially increase application latency and resource consumption. We develop a decompositional approach and a tool, called \sys, to systematically characterize the overhead of service meshes and to help developers quantify overhead in deployment scenarios of interest. Using \sys, we confirm that service meshes can have high overhead---up to 185\% higher latency and up to 92\% more virtual CPU cores for our benchmark applications---but the severity is intimately tied to how they are configured and the application workload. The primary contributors to overhead vary based on the configuration too. IPC (inter-process communication) and socket writes dominate when the service mesh operates as a TCP proxy, but protocol parsing dominates when it operates as an HTTP proxy. \sys also enables us to study the end-to-end impact of optimizations to service meshes. We show that not all seemingly-promising optimizations lead to a notable overhead reduction in realistic settings.

\section{Introduction}
Service meshes are fast becoming the {\em de facto} communication substrate for cloud applications. A survey of the Cloud Native Computing Foundation (CNCF) community found that 68\% of the organizations are already using or planning to use service meshes in the next 12 months~\cite{cncf-survey}. In-production use of service meshes has been growing 40-50\% annually~\cite{cncf-survey}.

Service meshes are popular because they solve important problems related to communication among loosely coupled (micro) services---the dominant paradigm for modern cloud applications~\cite{twitter-microservice,gan2019open,kakivaya2018service}. This includes discovering where services are located, establishing secure connections, and handling communication failures. They also offer many advanced capabilities such as rate limiting, load balancing, and telemetry, via built-in or custom message processing filters.

However, service meshes are not without downsides. A primary one is overhead. All application traffic traverses software proxies, called {\em sidecars}, which increases request latency and consumes more resources. Service meshes can add tens of milliseconds to request latency in some settings~\cite{linkerd-performance} and, they can consume multiple (virtual) CPU cores even at moderate load~\cite{istio-performance}. These overheads can degrade user experience, increase operational costs, and decrease revenue~\cite{kohavi2007online,flach2013reducing}.

Today, outside of a few point studies~\cite{linkerd-performance,istio-performance}, there is little systematic understanding of service mesh overheads. We do not know even basics such as the amount of overhead for real applications and what factors contribute most to the overhead. This is more than a matter of curiosity. Application developers do not know how much overhead the service mesh adds to their application, a problem exacerbated by the large configuration space of service meshes, each with different performance implications. Thus, they cannot evaluate functionality-performance trade-offs and judge the best way to configure the service mesh for their application. Further, there are several efforts in industry on lowering service mesh overhead~\cite{cillium-sm,istio-ebpf}. Without a proper accounting of the overhead and its major contributors, it is hard for these developers to quantify the effectiveness of their optimizations, especially as it relates to end-to-end impact on applications. 

Our goal is to characterize service mesh overhead in a way that helps application and service mesh developers reason about performance. To support these developers, unlike prior studies, we cannot simply measure the overhead of sidecars as a black box. If we did that, we would have to measure the combinatorial combination of all factors that impact overhead including service mesh configuration and application workload. And we still won't be able to judge the end-to-end impact of optimizing a specific aspect of service meshes.

We develop an approach that models the sidecar's operation as a composition of several independent components (e.g., read, write, and IPC) and key aspects of the workload. By characterizing individual components, we can estimate the overhead of a sidecar based on the components used in a given configuration.
Then, given workload characteristics such as the call graph, request rate, and message sizes, we can estimate the end-to-end overhead for the application (without the developer having to deploy the application under that configuration). Similarly, if an optimization reduces the overhead of some components, we can estimate its end-to-end impact. 

We build a tool called \sys that uses the approach above. We use it to quantify the latency and CPU overhead of Envoy~\cite{envoy}. It is the dominant sidecar, used by many service meshes~\cite{anthos,aspen,openshift,appmesh,consulsm,msopen}, including Istio~\cite{istio}, the most popular service mesh today~\cite{cncf-survey}. 

Using \sys, we conduct a systematic study of service mesh  overheads. We confirm that service meshes can have substantial performance penalty. Across two popular benchmark applications~\cite{gan2019open,onlineboutique}, depending on the configuration, request latency increases by 30-185\% and CPU usage increases by 41-92\%. In a large dataset of microservice-based applications~\cite{luo2021characterizing}, we find that using Envoy increases latency by up to 100\,ms and consumes 200 more virtual CPU cores for a quarter of the applications.  We also find that, for a given service mesh configuration, the overhead for different applications varies by multiple orders of magnitudes. Such high variation based on service mesh configuration and on application characteristics validate the need for a tool that developers can use for their specific deployment scenarios.

The compositional approach of \sys provides insight into the sources of overheads. We find that when configured in HTTP or gRPC mode, protocol parsing alone represents 62-73\% of the total overhead. In TCP mode, most overhead stems from inter-process communication (IPC) and socket write operations. The overhead of individual filters varies a lot. Some add as little as 3\% extra latency on top of the baseline overhead, while others add as much as 85\%. 

We also use \sys to evaluate the end-to-end overhead reduction for service meshes using two Linux kernel features: 1) Unix domain sockets in place of TCP connections for IPC, and 2) zero-copy writes for TCP sockets. Given that IPC and socket writes are substantial contributors to overhead, we wanted to understand if these features help reduce overhead. We model the performance of our components in the presence of these two features and evaluate their end-to-end impact on our microservices dataset~\cite{luo2021characterizing}.  We find that Unix domain sockets are helpful, reducing the average latency overheads by 27\% and the CPU overheads by 18\%, for TCP proxy. But zero-copy socket writes have negligible improvements. For small message sizes that are common to microservice workloads, the additional system calls in implementation negate the savings from avoiding the copy operation. 

We make three key contributions in our work:
\begin{itemize}
  \item A decompositional approach to model and predict service mesh overheads in specific deployment scenarios. It is based on our analysis of key components in service mesh datapaths.
  \item A tool that application developers can use it to quantify overhead and make judicious performance-functionality trade-offs. Service mesh developers can also use the tool to quantify the end-to-end impact of their optimizations.
  \item A systematic study of service mesh overheads. It confirms that the overhead can be significant in some settings and reveals which components contribute the most overhead in different settings. 
\end{itemize}

Our tool and findings will inform work on improving the performance and reducing the resource consumption of service meshes, which now play a central role in the modern application ecosystem.

\section{Background}
\label{sec:background}

\begin{figure}[t!]
\centering
\includegraphics[width=0.45\textwidth]{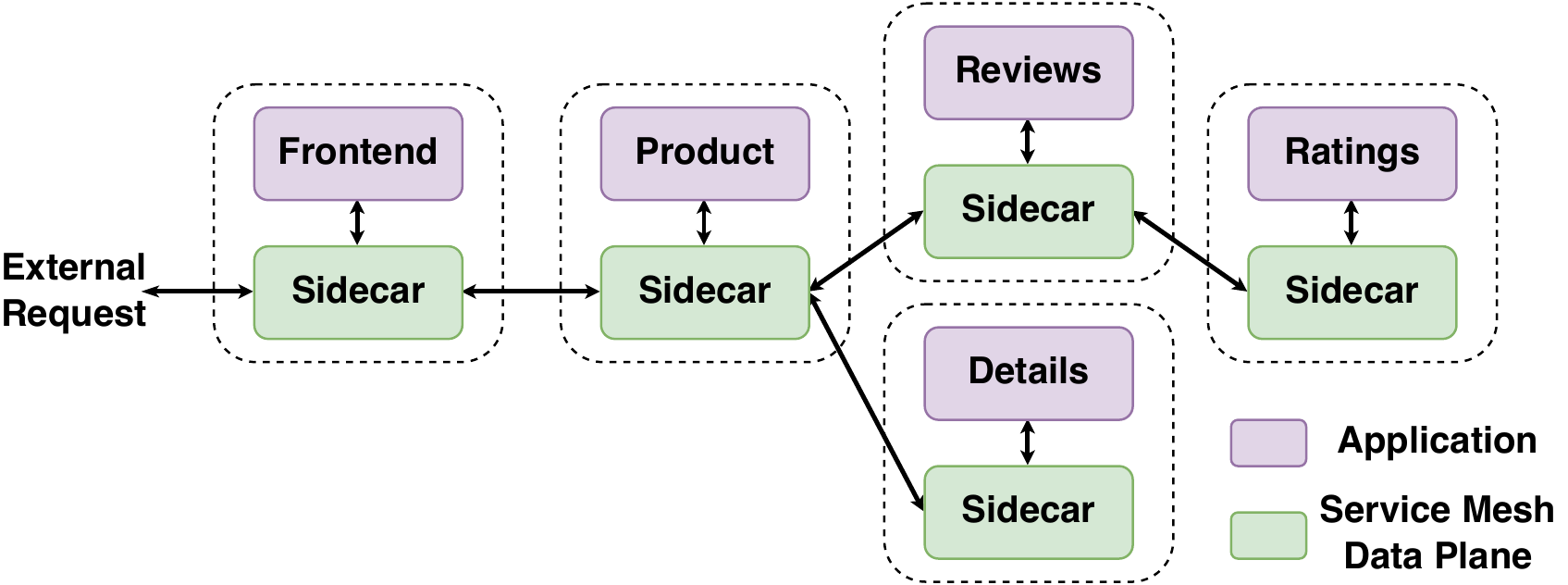}
\vspace{-2mm}
\caption{Bookinfo application with a service mesh. Image Redrawn from \cite{istio-bookinfo}.}
\vspace{-4mm}
\label{fig:mesh_arch}
\end{figure}

Service meshes emerged to solve a set of problems that arose when applications moved to the microservices architecture.
Instead of being a monolithic unit, applications are composed of multiple microservices. Figure~\ref{fig:mesh_arch} shows an example where the BookInfo application is composed of five individual services. Each microservice is run as an independent process (or container), often on different hosts, and can be scaled independently. 
Such decomposition enables agile application lifecycle management, fault-tolerance, scalability, and reuse of building blocks across applications~\cite{gheith2016ibm}. Modern applications commonly use the microservices architecture, with applications having tens of services~\cite{kakivaya2018service,twitter-microservice,gheith2016ibm}. 

While there are important benefits, splitting application into multiple microservices creates new problems as well. Developers must now figure out how services discover, communicate, and authenticate to each other. They must also figure out how to monitor and secure inter-service communication and how to handle failures. Early adopters of microservices built custom communication frameworks to solve these problems~\cite{hystrix,finagle}. Service meshes solve them in a reusable manner while also providing other functionality such as rate limiting and load balancing. 

The operation of service meshes is logically split into a control plane and a data plane. The control plane handles service discovery, metric collection, and certificate management, and it appropriately configures the data plane. The data plane uses software proxies called {\em sidecars}. A sidecar instance is co-located with each instance of an application service, which enables it to mediate all of service's network access to apply network policies, enforce encryption and log statistics. For example, in Figure~\ref{fig:mesh_arch}, to balance load across multiple instances of the Product service, the Frontend sidecar can spread Frontend-to-Product connections across different Product instances (multiple instances not shown in the figure). The service mesh control plane tracks where all the instances are running and configures the Frontend sidecar accordingly.

\begin{figure}[t]
\centering
\includegraphics[width=0.475\textwidth]{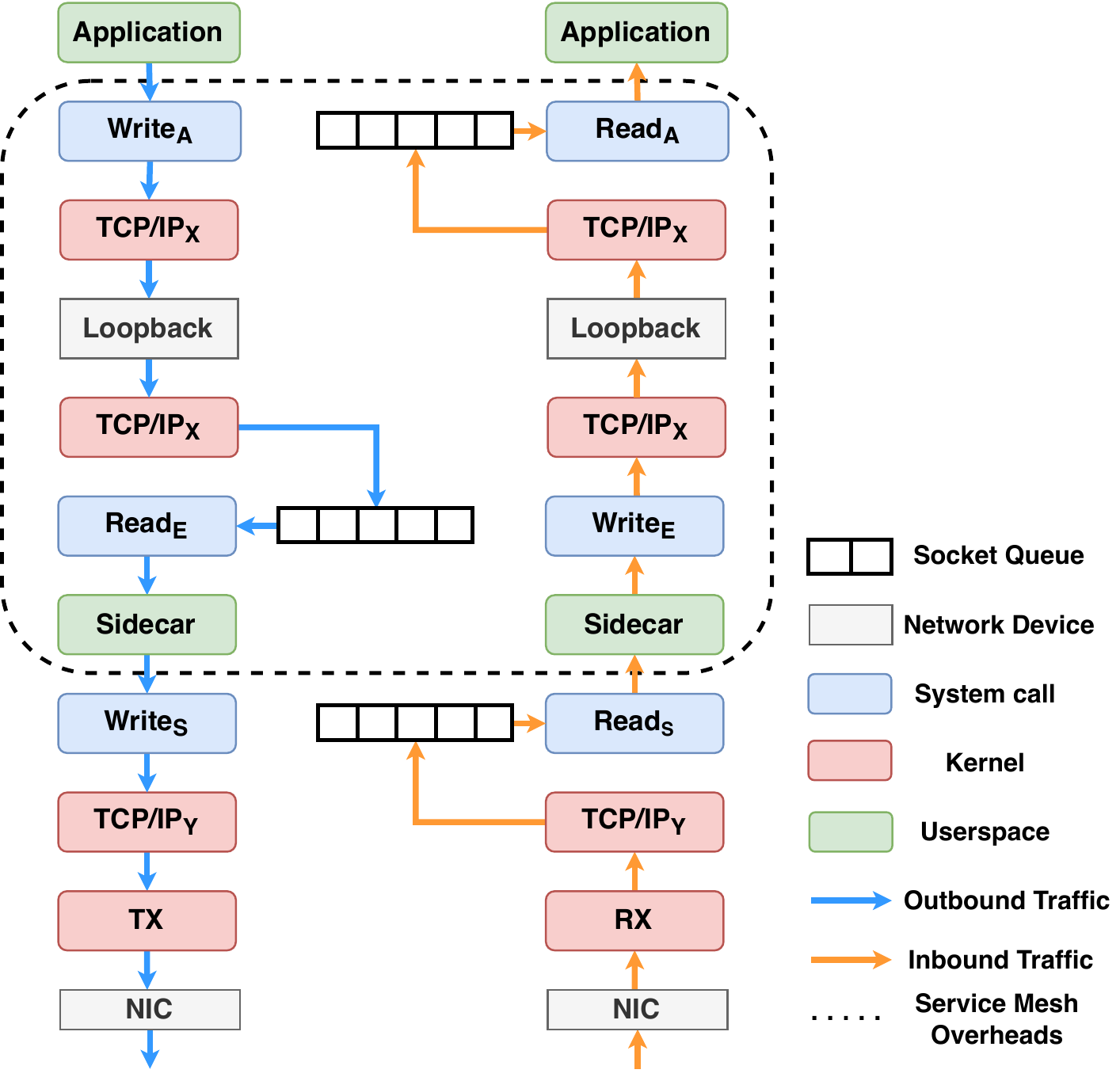}
\vspace{-2mm}
  \caption{Outbound (left) and Inbound (right) messages with a service mesh. Write\textsubscript{A}/Read\textsubscript{A} and Write\textsubscript{S}/Read\textsubscript{S} denote, respectively, the write/read of application and its sidecar. TCP/IP\textsubscript{X} denotes the TCP/IP stack of network namespace X. The figure assumes that the application uses the same event notification interface (i.e., epoll). The extra steps added by the service mesh are in the dashed box.}
\vspace{-4mm}
\label{fig:data_path}
\end{figure}

\parab{Service Mesh Data Path}
\label{sec:data-path}
The advantages of service meshes come at the cost of performance and resource overhead. We focus on the data plane overhead as it impacts every request and is on the critical path of user experience. We use Envoy~\cite{envoy} as our example. It is used by many service meshes~\cite{anthos,aspen,openshift,appmesh,consulsm,msopen}, including Istio, the most popular service mesh. Other data plane proxies~\cite{linkerd} have a similar architecture.

\autoref{fig:data_path} shows data path for both outbound and inbound traffic. During initialization, the control plane adds iptable rules that redirect all inbound and outbound traffic to the sidecar. As a result, logical connections between microservices are broken into three separate connections: two connections between the sidecars and their associated microservices, and one connection between the sidecars. When the application sends a message, it executes a write system call. The kernel network stack and the loopback device process the message and notify the sidecar. The sidecar then reads the data from the kernel, processes it, and writes it back to the kernel. Finally, the NIC driver transmits the message. Similar to the inbound traffic, upon receiving a message, the sidecar intercepts the message before it is passed to the application.

\begin{figure}[t]
\centering
\includegraphics[width=0.46\textwidth]{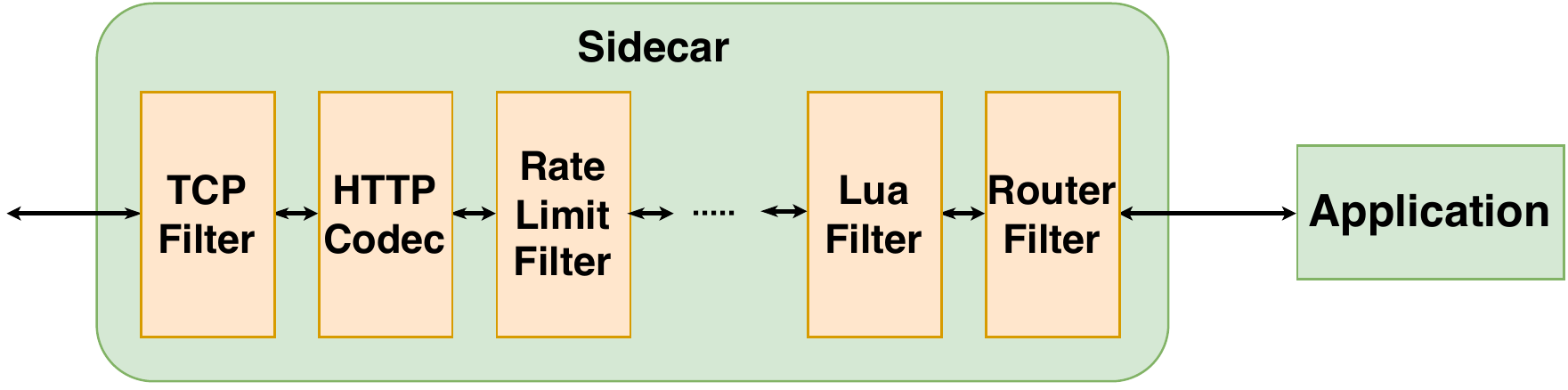}
\vspace{-2mm}
\caption{Message processing inside a sidecar.}
\label{fig:proxy_arch}
\end{figure}

The exact processing done by the sidecar depends on its configuration; Figure~\ref{fig:proxy_arch} shows the general data flow. When a message arrives, it is parsed based on the protocol of choice (e.g., TCP, HTTP,  gRPC). In TCP mode, traffic is treated as an opaque TCP stream; in HTTP and gRPC mode, messages are parsed as per the protocol which enables additional functionality specific to the protocol. After the message is parsed, it is processed by one or more {\em filters}. Filters are short programs that process individual messages and implement functions like traffic monitoring, rate limiting and fault injection. Envoy filters can be written using 1) C++ code, 2) Lua scripts, or 3) WebAssembly modules. All of Envoy's 40+ built-in filters are C++-based, while application developers tend to write custom filters using Lua or WebAssembly.

Given the data paths in Figures~\ref{fig:data_path} and~\ref{fig:proxy_arch}, we can see a number of sources of  overhead. First, without a sidecar, to send a buffer, the kernel copies the application buffer into a kernel buffer, which the NIC can subsequently access through Direct Memory Access (DMA). With a sidecar, the buffer must additionally be copied to the sidecar buffer and then back into a kernel buffer (resulting in two extra copies). Second, there are many additional system call invocations, such as the sidecar waiting for data through epoll and reading and writing buffer from/to kernel. Finally, using sidecars incurs extra IPC invocations (e.g., the loopback interface in Istio). In addition, sidecars may need excessive computation on the buffer, including parsing the data stream into data structures for HTTP, JSON, and RPC data formats.

\section{Modeling Service Mesh Overhead}

Our goal is to characterize the performance overheads of service meshes for real-world microservice applications. We want to support two classes of developers. First, application developers who are looking to deploy service meshes should be able to understand overhead as a function of  service mesh configuration (e.g., proxy types, filters), so they can appropriately trade off functionality and overhead. Second, service mesh developers should be able to understand the impact of their optimizations on real-world applications. Today, there are several ongoing directions in industry and academia to reduce the service mesh performance overheads~\cite{cillium, cui2021smartnic}. However, these service mesh developers do not have a way to get these insights (without doing their own extensive benchmarking experiments).

A key challenge we face toward meeting our goals is the large operational space of service meshes. An application may be running Envoy in one of many possible ways, each with different performance implications. There are at least three dimensions of variations: $i)$ type of proxy (e.g., TCP, HTTP, gRPC, and MySQL); $ii)$ which filters are used; $iii)$ application workload, where the salient characteristics are message sizes and rates. These variations (and their combinatorial combinations) mean that a black-box approach to measuring overhead is a non-starter. 

We must instead model the overhead of finer-grained components and then be able to compose the individual component overheads to predict overhead for a given deployment scenario. Our models are a concise representation of performance overheads over the entire operation space, including both the service mesh configuration and application workloads. This modeling approach allows us to reason about a service mesh's large operation space. For example, for an application developer to choose a appropriate configuration for a service mesh, instead of benchmarking every possible configuration, we can use our models to quickly predict the performance overheads for \textit{any} service mesh configuration for his/her workloads. However, this modeling-based approach requires us to carefully choose the granularity and nature of the components. If they are too fine-grained, accurately characterizing their overhead (and composing them) will be difficult; if they are too coarse-grained, we'll suffer from the same challenge as with black-box measurements.

\begin{table}[t!]
\begin{center}
\begin{tabular}{|l|p{0.14\textwidth}|p{0.23\textwidth}|} 
 \hline
 & Component & Description  \\ [0.5ex] 
 \hline\hline
1 & IPC & Data transfer between sidecar and application  \\ 
 \hline
2 &  Read & Read syscall and data copy from kernel to user space   \\
 \hline
3 &  Write & Write syscall, data copy from user to kernel space, and network stack's TX processing \\
 \hline
4 &  Notification & I/O event notification processing \\
 \hline
5 &  Protocol Parsing & Protocol parsing in sidecar \\ 
 \hline
6 &  Other Userspace & Other userspace processing in sidecar \\ 
 \hline
7 & Filter & Filter chain processing in sidecar \\ 
 [1ex] 
 \hline
\end{tabular}
\caption{Components in \sys's performance model. 
}
\label{table:model} 
\end{center}
\end{table}

Table~\ref{table:model} shows the components we profile. The first four represent the sidecar's interactions with the application and the kernel:
1) inter-process communication (e.g., loopback) between the application and its sidecar proxy, 2-3) writes and
reads from the sidecar proxy to the kernel, which also include the data copies (note that write also includes the TX's TCP/IP processing), and 4) blocking waits on socket ready notifications (e.g., from \texttt{epoll}), The last three breakdown message processing inside the sidecar: 5) parsing the messaging protocol (e.g., HTTP), 6) additional userspace processing in the sidecar (e.g., default Router filter), and 7) processing done by user-configured filters.

We do not claim that this is the only way to decompose the sidecar's overhead. We chose these components to balance the granularity concern above and because they are mostly independent. This independence allows us to easily compose their overheads to estimate total sidecar overhead. It also means that service mesh optimizations typically impact one or two of these components, which allows us to understand and predict the performance impact through targeted modifications to the profiles of optimized components. We have also chosen to ignore certain sources of overheads such as cache contention and more frequent context switching. Our experiments confirm that the impact of such factors is minimal.

\begin{figure*}[t]
\centering
\includegraphics[width=0.9\textwidth]{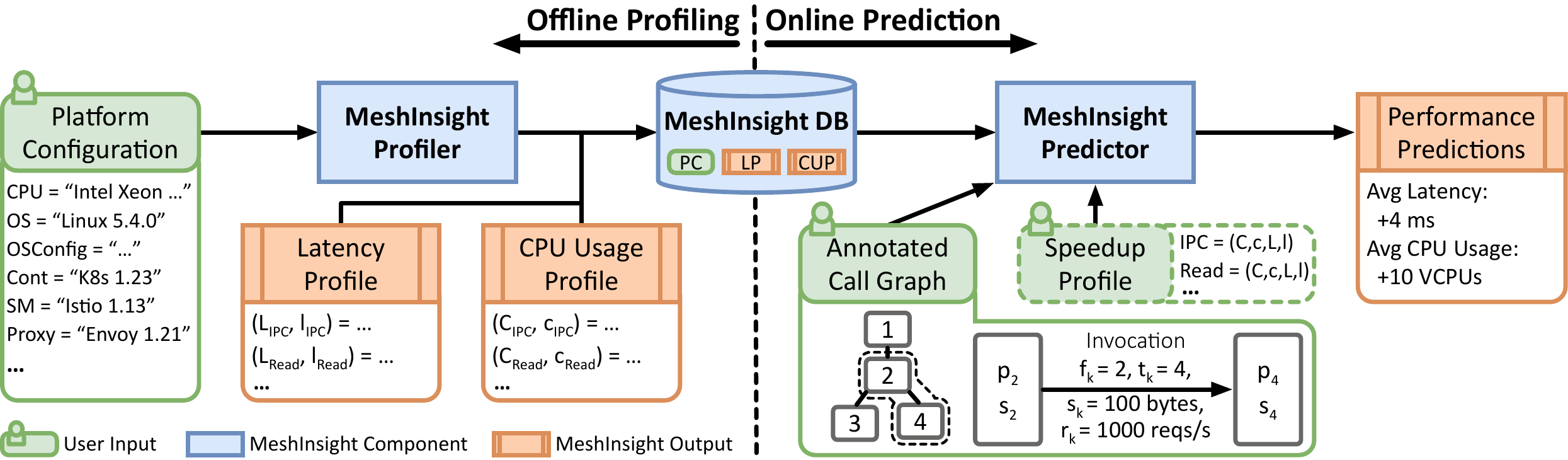}
\caption{Overview of \sys workflow. The \sys DB stores performance profiles associated with a hardware and software platform configuration, which are generated by the \sys Profiler during the offline profiling stage \autoref{sec:offline}. In the online phase \autoref{sec:prediction}, these profiles are used by the \sys Predictor, in conjunction with an annotated call graph (ACG) provided by the user, to compute latency and CPU performance predictions for an application deployment. The user can optionally provide a speedup profile, which \sys uses to adjust the predictions accordingly.}
\vspace{-2mm}
\label{fig:tool}
\end{figure*}

\parab{\sys overview}
\autoref{fig:tool} shows the workflow of \sys. It has an offline profiling phase and an online prediction phase. The offline phase generates performance profiles of individual service mesh components, and the online phase predicts overhead based on these profiles, service mesh configuration, and application workload.

The performance of a component is modeled as a function of message size and request rate because these two workload properties are the primary determiners of performance. In our experience, simple linear functions of these two properties suffice (see next section for details). The profile of a component is specific to the {\em platform}, which includes the hardware (e.g., CPU, memory), OS, and Envoy version. Overhead predictions are made only for previously profiled platforms. We make no attempt to predict performance for unprofiled platforms. \footnote{As part of this work, we are developing a shared repository of performance profiles for common platforms such as AWS instances, to enable reuse of profiling data.}

The online prediction phase uses performance profiles from the offline phase to provide performance predictions in the context of a specific application deployment scenario. These predictions are based on an annotated call graph (described in the next section), provided by the user. The annotated call graph encodes the infrastructure and service mesh configuration of each microservice and captures the relevant properties of a microservice application that is handling a particular request.

\parab{Overhead measures}
We consider two primary measures of interest to developers: latency and CPU usage overhead. In particular, we quantify extra latency service mesh adds to the application messages, and extra CPU (number of virtual cores) that message processing consumes. For latency, we consider the average overhead in a lightly or moderately loaded system. Latency-sensitive applications are more likely to be run in this regime. In heavily loaded systems, unpredictable effects due to resource contention kick in, which we do not model. We leave the study of overhead measures such as memory and tail latency to future work. 

\section{\sys Design}
\label{sec:tool}

We now describe the design of \sys in more detail.

\subsection{Building Component Profiles}
\label{sec:offline}

In the offline phase, for each component in Table~\ref{table:model}, \sys builds performance profiles that characterize components' message processing latency and CPU usage as a function of message size and rate. To build these profiles, we exercise these components under a few different settings and then interpolate and extrapolate their performance to settings that are not directly measured. 

We conduct the following types of profiling runs: $i)$ sidecar configured as TCP, HTTP, or gRPC proxy, with no additional filters; and $ii)$ sidecar configured with only the filter(s) of interest. Profiling runs use an echo server paired with a sidecar. We use wrk\cite{wrk} and wrk2~\cite{wrk2} both as load generators as tools to measure the end-to-end latency with high precision. To mimic a lightly loaded server, the wrk client generates requests such that at most one request is outstanding at any time. 

To quantify the overhead of individual components during a run, we exploit the fact that all components (except filters--see below) corresponds to a specific kernel- or user-space function. We measure the latency of each component using a modified version of BCC's funclatency~\cite{funclatency}, an eBPF-based tool that uses Kprobe and Uprobe to monitor time spent on a function. We measure CPU usage of each component using the standard sampling technique~\cite{perf}, which allows us to quantify the CPU consumption of any function. We identify the function for a component using a mix of function name, function's input/output and process ID.

Filters present two wrinkles in this process. First, most of them do not have a known function. We measure the overhead of a filter by subtracting the overhead of a setup without the filter from an otherwise identical setup with it.  Second, the overhead of some filters depends significantly on certain configuration parameters. For instance, the overhead of the Rate Limit filter, which limits the network traffic to a service, depends on whether the developer needs limit traffic rate on the entire service mesh (global rate limiting) or on per service instance basis (local rate limiting). Likewise, the overhead of Tap filter, which logs traffic, depends on whether the log is written to a file or sent over network. In our profiling, we treat such filters as different components.

Data from the profiling runs and the following assumptions (empirically validated in \S\ref{sec:characterization}) enable us to estimate the sidecar's overhead in any setting.

\begin{itemize}
    \item \textbf{A1:} The total latency and CPU overhead of the sidecar is the sum of components' overhead.
    \item \textbf{A2}: Latency overhead is a linear function (see below) of message size. 
    \item \textbf{A3}: CPU overhead is a linear function of message size and proportional to message rate. 
\end{itemize}

Thus, to estimate the overhead of the sidecar in a configuration with multiple filters, which has not been measured directly, we can add the overhead of the base configuration without filters and the overheads of filters that are employed.

To estimate the latency of a component $x$ for a message size $s$, per prior work~\cite{li2019socketdirect,stewart2005performance}, we use this linear function: 
\begin{align}
L_x + s \times l_x
\label{eq:latency_model}
\end{align}
where $L_x$ is the baseline message processing latency and $l_x$ is the per-byte processing latency. We assume that components' per-message latency does not vary based on request rate. 

To estimate CPU consumption of a component $x$ for request rate $r$, we use the following linear function:
\begin{align}
   r \times (C_x +  s \times c_x)
\label{eq:cpu_model}
\end{align}
where $C_x$ denotes the baseline per-message CPU usage and $c_x$ denotes the per-byte CPU usage. 

The impact of message size captured in the two equations above assume that a message is processed by each component (e.g., read or written) as one unit. This assumption may be violated for large message sizes when they are split into multiple units.  Size threshold at which a message is split may be overridden by applications or Envoy, but is typically at least a few KB; it was always above 4KB for platforms that we have experimented with. The implication for our modeling is that it will underestimate the overhead for messages that are split; the actual latency and CPU cost is higher for such messages. 
We quantify this underestimation in \autoref{sec:size_and_rate}.
Fortunately, the vast majority of messages sizes are small~\cite{qian2009tcp,montazeri2018homa,yang2020large}, and the impact of our modeling approximation is therefore minimal.

To estimate $(L_x, l_x)$ and $(C_x, c_x)$, \sys profiles component for five different message sizes (i.e., 100B, 1KB, 2KB, 3KB, 4KB) and uses linear regression on the resulting data. These two tuples represent the latency and CPU profile of a component for a particular platform. 

\subsection{Predicting Overhead}
\label{sec:prediction}

Application developers can use \sys to estimate service mesh overhead in any deployment scenario of interest by providing an {\em  annotated call graph} (ACG). The ACG captures the details of the deployment and interactions among microservices in response to a request. 

Formally, an ACG is a tuple $(V, P, S, G)$, where $V$ = \{$v_1$, $v_2$, ..., $v_n$\} is a set of vertices representing microservice instances; $P$ is a map from microservice instances to platforms; and $S$ is map from microservices instances to configurations (i.e., protocol and filters). $G$ is a DAG (directly acyclic graph) of microservice invocations. Each node in the graph is an invocation, and edge represents invoked-after relationship. An invocation $k$ is a tuple $(f_k, t_k, s_k, r_k)$, where $f_k$ is the calling microservice (empty if called externally), $t_k$ is the microservice invoked, and $s_k$, $r_k$ are the expected size (in bytes) and rate of messages (in requests/second) along this invocation.

\begin{figure}[t]
  \centering
   {\includegraphics[width=0.9\linewidth]{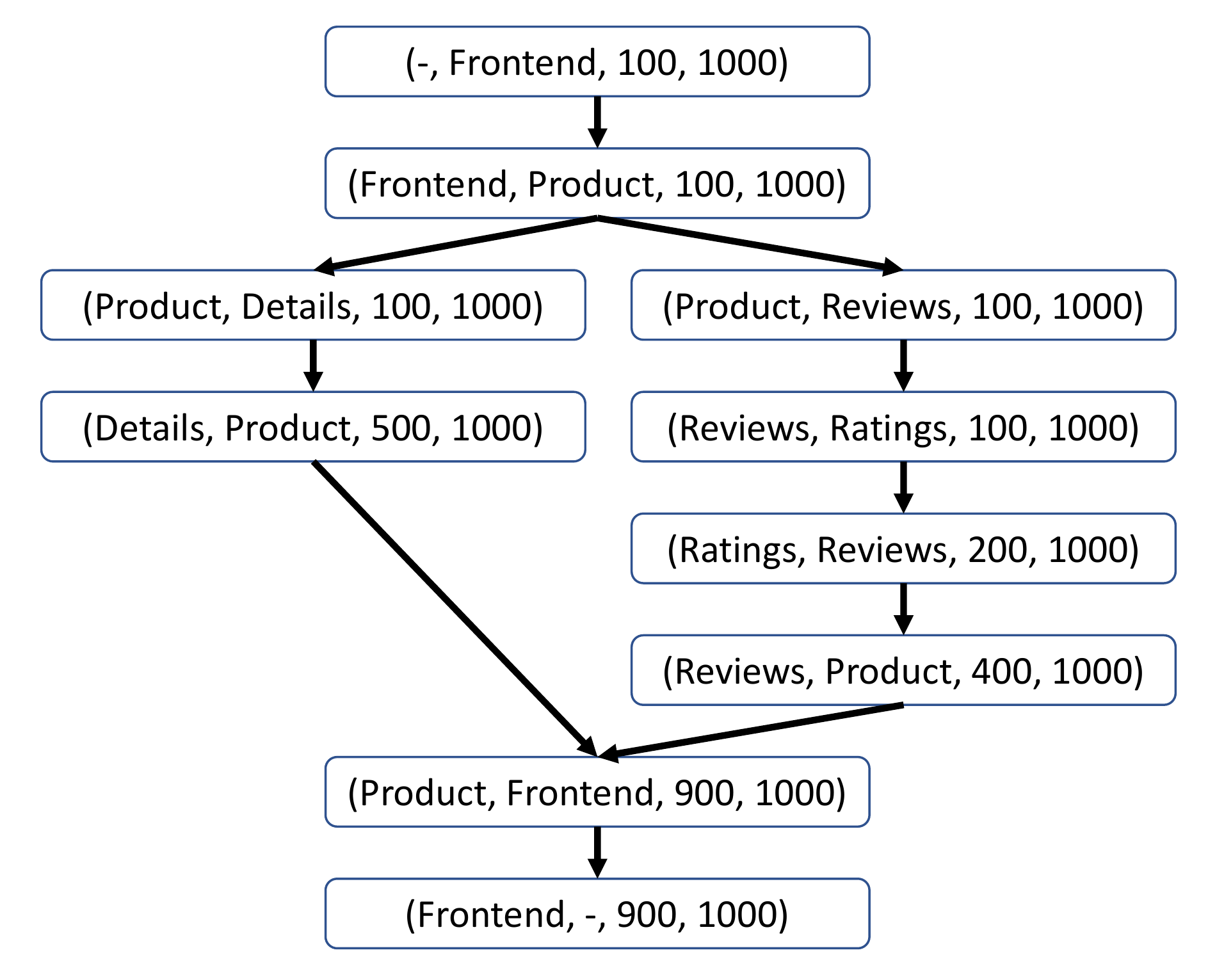}}
  \caption{An example call graph for the Bookinfo application.}
  \label{fig:example-g}
\end{figure}

\autoref{fig:example-g} show an example of $G$ for the Bookinfo application in \autoref{fig:mesh_arch}. An external client calls Frontend, which in turn calls Product. The Product service calls Reviews and Details in parallel. Reviews calls Ratings and responds to Product after getting the response. Product responds to Frontend after both Details and Reviews respond. Finally, Frontend responds to the external client. 

\sys estimates the overhead of a given ACG. We expect developers have the information necessary to furnish the ACG. If an application has multiple request types (as is the case for our example applications in \S\ref{sec:benchmarks}), developers can provide multiple ACGs and learn the overhead of each. In cases where an invocation is non-deterministic (e.g., based on cache hits), developers can provide two ACGs, one with the invocation and one without. They will then learn the overhead of each ACG and combine them based on expected probabilities to get the average overhead. Similarly, if a service load balances to multiple instances of a downstream service, the developers can provide multiple ACGs, one for each possible path through the services. In the Bookinfo example, if Frontend balances load across two Product instances, we will have two ACGs.  The message rate from Frontend to each Product instance, and downstream of that, will be 500 requests per second.

\parab{Generating Predictions.}
Given an ACG, \sys estimates the latency and CPU overhead. It starts by computing the overhead of each invocation. For an invocation $(f_k, t_k, s_k, r_k)$, the overhead is based on messages of the given size and rate leaving the service at $f_k$ and entering the service at $t_k$. The sidecar configuration for these services tell us which components are exercised. We compute the component-level overhead (using $s_k$ and $r_k$) and then sidecar-level overhead by summing component overheads. In some configurations, not all components are subjected to the same $s_k$ and $r_k$. For instance, if a fault injection filter, which can be configured to drop some messages, is present, downstream filters will see a lower message rate. Currently, we ignore such intra-sidecar variations, though it is straightforward to extend our model to account for them.

\sys computes end-to-end request-level overheads using sidecar-level overheads as computed above. For CPU, this is simply summing all the sidecar-level overheads. For latency, simple summation does not work because computations can happen in parallel, and thus critical path analysis is needed. In the Bookinfo example above, the end-to-end latency depends on which of Details or Reviews is slower to respond to Product, and the latency of the faster one is not critical. We thus compute the latency overhead using critical path analysis, essentially reporting the highest latency path in G. In this analysis, we are assuming that the critical path computed based on latency overheads is the same as the one with application processing; this path will often be the longest path in the invocation chain. If information on application processing latency were available, \sys could compute a better estimate of end-to-end latency overhead in cases where this assumption does not hold. In the future, we will allow developers to optionally provide this information.

\parab{Quantifying the impact of service mesh optimizations.}
The prediction techniques described above can be used by service mesh developers to estimate the end-to-end impact of their optimizations. To enable this estimation, service mesh developers need to provide information on the impact of their optimization for the component(s) they have optimized. That is, they need to update the performance profiles. This update may be based on the estimated impact of their planned optimization (which has not been implemented yet). For example, the developer may estimate that their optimization will lower the baseline write overhead by 50\%. Alternatively, new performance profiles may be based on running \sys profiling after implementing the optimization. 

Once information on new performance profiles is provided, \sys can estimate the overhead of the new system and how much improvements is bring compared to the original.

\section{Characterizing Service Mesh Overhead}
\label{sec:characterization}

We now use \sys to characterize the overhead of service meshes in realistic deployments scenarios and shed light on major sources of overhead in different scenarios. Our experiments use Cloudlab~\cite{cloudlab} machines with two 16-core Intel Xeon Gold 6142 CPUs (2.6 GHz) and 384GB RAM, Ubuntu 20.04 LTS (Linux kernel v5.4.0), Kubernetes v1.12.5, Istio v1.13.0, and Envoy 1.21.0. We disable TurboBoost, CPU C-states and dynamic CPU frequency scaling to reduce measurement variance. 

\subsection{Application Benchmarks}
\label{sec:benchmarks}

\begin{figure}[t!]
\centering
\includegraphics[width=0.42\textwidth]{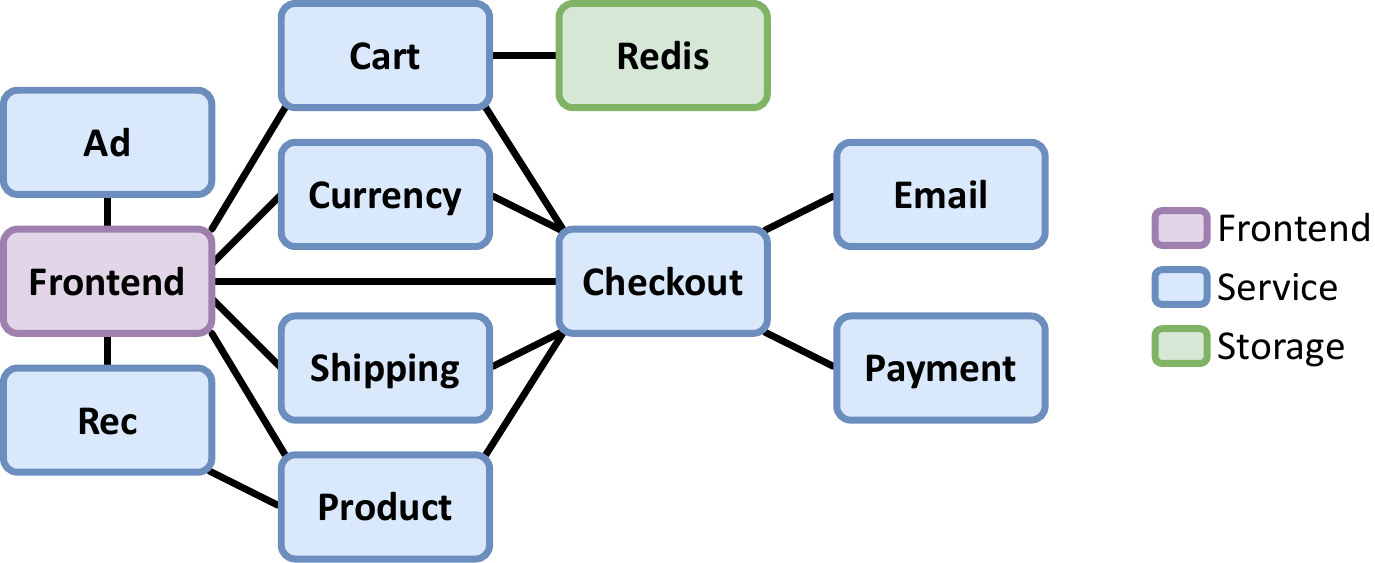}
\vspace{-2mm}
\caption{Online Boutique application \cite{onlineboutique}.}
\vspace{-2mm}
\label{fig:boutique_arch}
\end{figure}

\begin{figure}[t!]
\centering
\includegraphics[width=0.475\textwidth]{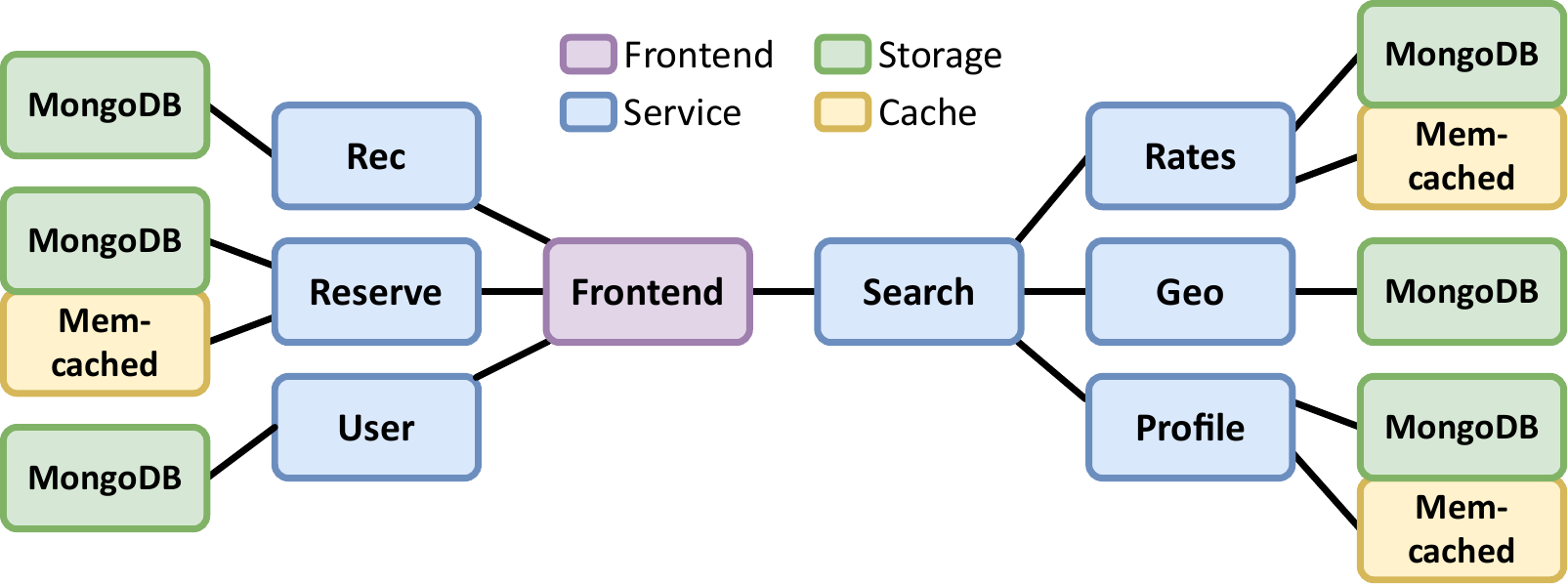}
\vspace{-2mm}
\caption{Hotel Reservation application \cite{gan2019open}.}
\vspace{-4mm}
\label{fig:hotel_arch}
\end{figure}

\begin{figure*}[t!]
  \centering
    \subfigure[Latency. \label{fig:benchmark_latency}]{\includegraphics[width=0.5\linewidth]{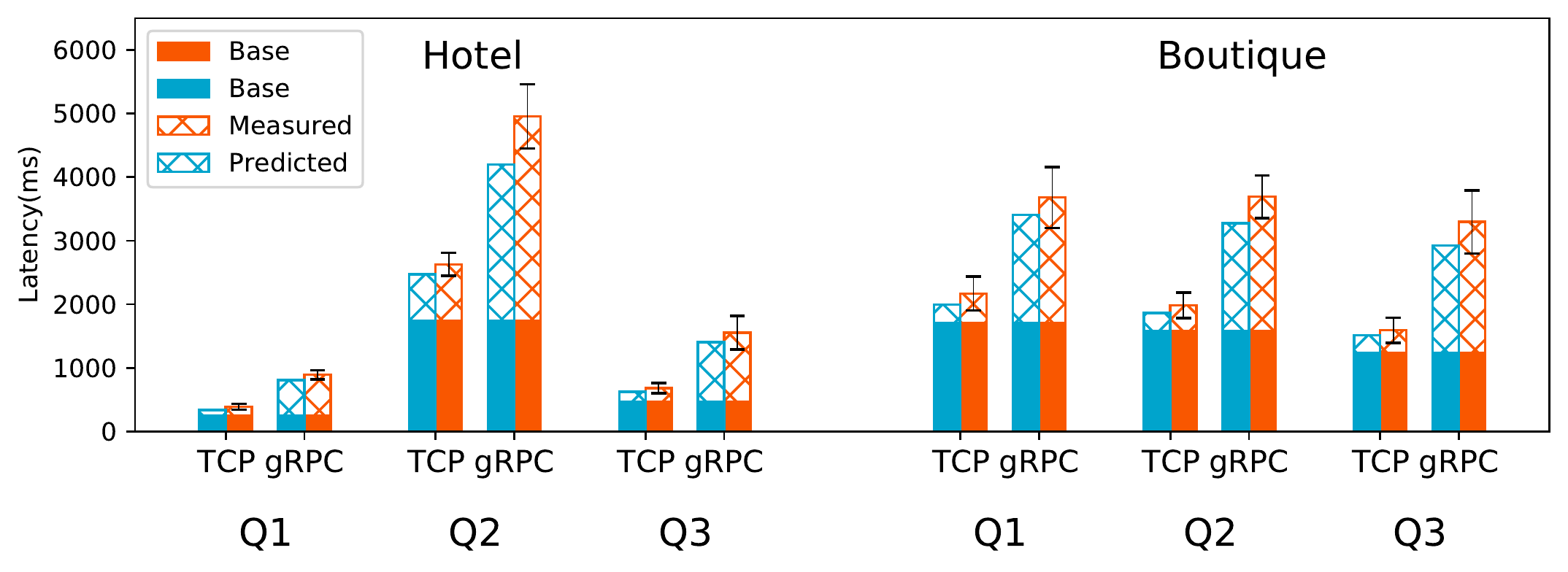}}\hfil 
    \subfigure[CPU Usage. \label{fig:benchmark_cpu}]{\includegraphics[width=0.5\linewidth]{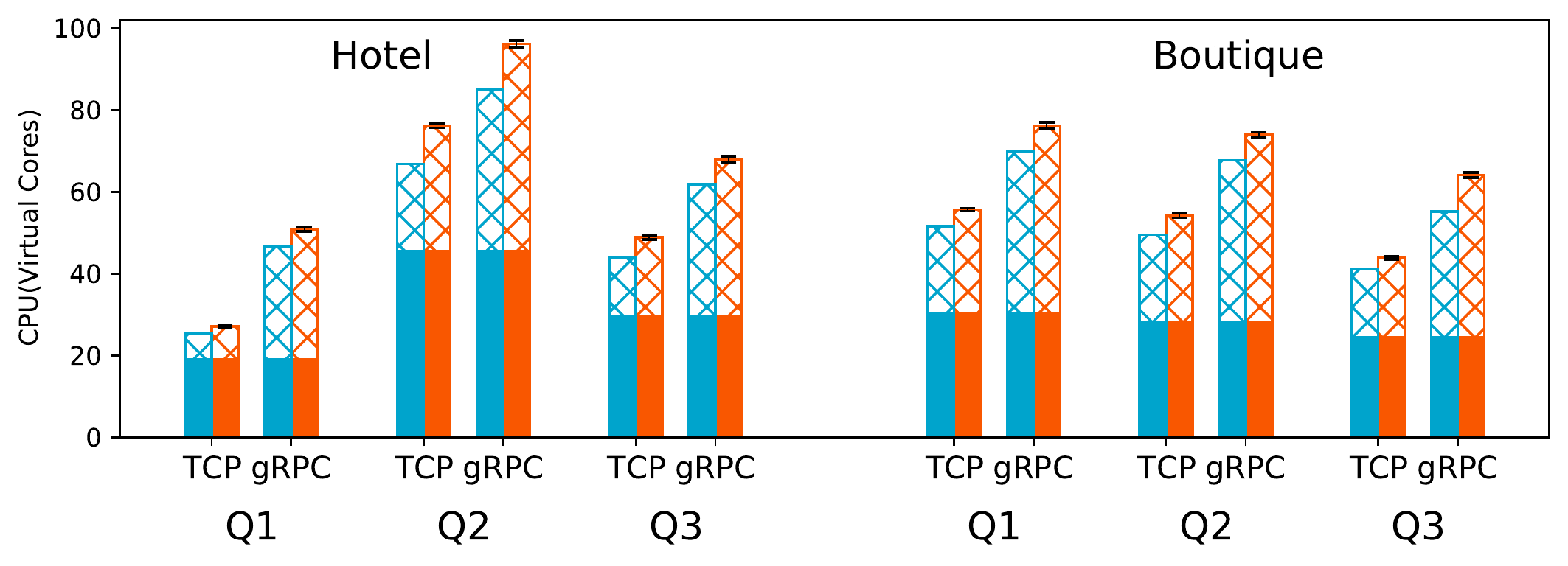}}
  \vspace{-2mm}
  \caption{Predicted and measured overhead for Online Boutique and Hotel Reservation applications. Base denotes latency and CPU usage when the application is run without a service mesh. The error bars for measured overhead are standard deviations.} 
  \label{fig:benchmark}
\end{figure*}

To characterize the overhead of service meshes on realistic applications, we consider two popular microservices benchmarks: Online Boutique\cite{onlineboutique} and Hotel Reservation\cite{gan2019open}. Online Boutique (\autoref{fig:boutique_arch}) has 11 microservices. It is a web-based e-commerce application where users can browse items, add them to the cart, and purchase them. Microservices are written in different languages (Python, C\#, Java, and Go) that communicate using gRPC. Hotel Reservation (\autoref{fig:hotel_arch}) has 17 microservices and supports searching for hotels using geolocation, making reservations, and providing hotel recommendations. All microservices are implemented using Go and communicate using gRPC. 

We deploy these applications on multiple Cloudlab hosts and consider two deployment scenarios: TCP and gRPC. (HTTP cannot be used because the applications are gRPC-based.) In TCP mode, all sidecars are configured as a TCP proxy that relays the message to the application service. In gRPC mode, the sidecars parse the gRPC stream and collect basic application-level metrics. 

We consider three queries for each benchmark. For hotel reservation, the queries are:
\begin{enumerate}
    \item \textbf{(Q1) User}: Checks the username and password. Calls User and its MongoDB.
    \item \textbf{(Q2) Search}: Returns available hotels based on location and check-in/check-out dates. Calls Search, Rates, Geo, Profile, Reserve
    \item \textbf{(Q3) Reservation}: Reserves a hotel room. Calls Reserve, User and their MongoDB and Memcached storage.
 and their MongoDB and Memcached storage.
\end{enumerate}

For online boutique, the queries are: 
\begin{enumerate}
    \item \textbf{(Q1) Index}: Returns the home page. Calls Currency, Products, Cart and Ad.
    \item \textbf{(Q2) Browse\_Product}: Returns product details. Calls Product, Currency (twice), and Cart.
    \item \textbf{(Q3) View\_Cart}: Returns the user's shopping cart. Calls Cart, Shipping, Product, and Currency.
\end{enumerate}

We derive the ACG (annotated call graph) for each query and force a cache miss on each memcached access. The message sizes in each ACG are based on the actual traffic of the application; we find that most messages are small (few hundred bytes). We set the request rates close to the maximum the machine can sustain in gRPC mode for each query. 

\tabcolsep=0.4cm
\begin{table*}[t]
\small
\begin{tabular}{cr@{\hskip 4pt}lr@{\hskip 4pt}lr@{\hskip 4pt}lr@{\hskip 4pt}lr@{\hskip 4pt}lr@{\hskip 4pt}l}
\toprule
 & \multicolumn{6}{c}{\textbf{Latency (us)}} & \multicolumn{6}{c}{\textbf{CPU Usage (Virtual Cores)}} \\
\cmidrule{2-7} \cmidrule{8-13}
 & \multicolumn{2}{c}{\textbf{TCP}} & \multicolumn{2}{c}{\textbf{HTTP}} & \multicolumn{2}{c}{\textbf{gRPC}} & \multicolumn{2}{c}{\textbf{TCP}} & \multicolumn{2}{c}{\textbf{HTTP}} & \multicolumn{2}{c}{\textbf{gRPC}} \\
\midrule
\textbf{IPC} & 11.59 & (30\%) & 12.75 & (8\%) & 13.04 & (7\%) & 0.49 & (15\%) & 0.51 & (5\%) & 0.55 & (4\%)\\
\textbf{Read} & 8.14 & (16\%) & 9.01 & (5\%) & 9.37 & (5\%)& 0.26 & (8\%) & 0.29 & (3\%) & 0.30 & (2\%)\\
\textbf{Write} & 13.22 & (34\%) & 13.80 & (8\%) & 14.35 & (7\%) & 0.45 & (14\%) & 0.48 & (5\%) & 0.57 & (4\%)\\
\textbf{Notification} & 1.33 & (3\%) & 1.27 & (1\%) & 1.35 & (1\%) & 0.26 & (8\%) & 0.27 & (3\%) & 0.26 & (2\%)\\
\textbf{Protocol Parsing} & - & & 117.35 & (70\%) & 142.38 & (73\%) & - & & 6.00 & (62\%) & 9.76 & (71\%)\\
\textbf{Protocol Other} & 4.25 & (11\%) & 13.07 & (8\%) & 14.39 & (7\%) & 1.79 & (55\%) & 2.09 & (22\%) &  2.34 & (17\%)\\
\midrule
\textbf{Total} & 38.63 & & 167.25 & & 194.79 & & 3.25 & & 9.65 & & 13.79\\
\bottomrule
\end{tabular}
\vspace{2mm}
\caption{Contribution of different components to the overhead of a single sidecar instance in different protocol modes. The numbers report both inbound and outbound overheads.}
\vspace{-3mm}
\label{tab:breakdown-100B}
\end{table*}

\subsection{End-to-end Overhead}
\label{sec:characterization-e2e}

\autoref{fig:benchmark} shows the base (without the service mesh) latency and CPU usage of the application along with predicted and measured overheads. The measured overhead is the latency (or CPU usage) of running the application with the service mesh minus that of running it without the service mesh.

We see that service meshes can be a significant source of overheads. When operating in gRPC mode, it can increase latency by up to 185\% and consume 92\% more CPU. In TCP mode, the overhead is lower but still substantial--latency increases by 22\% and CPU usage by 43\%. The next section sheds light on why these two modes behave differently.

Service mesh overhead will increase further as filters are added (see below). These high overheads, and differences in overheads in different settings, is why \sys is needed to enable developers to appropriately trade-off performance and functionality. For instance, TCP mode might be enough for most services, and gRPC mode is limited only to services where extra control or visibility is required.

In \autoref{fig:benchmark}, we can also see that \sys predictions track well the overhead for all queries across both benchmarks even though their individual performance varies significantly. The predictions are generally on the lower end of measured values.  
This underestimation stems from random noise in complex, running systems and \sys ignoring components with minimal overheads (e.g., kernel scheduling). 

\subsection{Sources of Overhead}
\label{sec:characterization-breakdown}

To shed light on sources of overhead, \autoref{tab:breakdown-100B} shows how much each component contributes when Envoy is run in different protocol modes and without any filters. This experiment uses a synthetic application (echo server) to which a sidecar can be attached in TCP, HTTP, or gRPC modes. It uses 100 byte messages at 30K requests per second. 

We can draw several conclusions from this data. First, using HTTP and gRPC is substantially more expensive than TCP. The additional overhead of HTTP is roughly 4x for latency and 3x for CPU; for gRPC it is 5x for latency and 4x for CPU. The bulk of this additional overhead stems from protocol parsing, accounting for 62-73\% of the total overhead. 

Parsing overhead is unfortunate because the application code will spend resources on parsing as well. Because of the way in service mesh data path is organized today, there is no opportunity to reuse parsing work across Envoy and the application. If enabled, such reuse will have a notable impact on the architecture of service meshes.

Second, we see that IPC overhead is notable (34\% for TCP) and notification overhead is small (3\% for TCP). This observation implies that asynchronous processing between the application and sidecar is not expensive by itself, but the default IPC mechanism in Envoy is expensive. We can tackle this overhead by either putting the sidecar in the same process as the application. However, this can have security implications because a malicious application may circumvent the network policies. In addition, upgrading a sidecar more complicated would require recompiling the application. Another option is to use a more lightweight IPC mechanism. We will consider the second option in the next section.

Third, some components have disparate impacts on latency and CPU. This disparity most pronounced for "Protocol Other", where its contribution to CPU overhead is far greater than its contribution to latency, but hold for other components as well (e.g., Write). It implies that some optimizations may impact one type of overhead and not the other, and developers need to be careful that optimizing for latency does not hurt CPU and vice versa.

\subsection{Impact of Filters}
\label{sec:filters}

We now characterize the impact of filters on overhead. We find that filters can be quite expensive (even configured as a no-op) and, as assumed by \sys, validate that their overhead is additive.

\tabcolsep=0.3cm
\begin{table}[t]
\begin{tabular}{cr@{\hskip 4pt}lr@{\hskip 4pt}l}
\toprule
 & \multicolumn{2}{c}{\textbf{Latency(us)}} & \multicolumn{2}{c}{\textbf{Virtual Cores}} \\
\midrule
\textbf{Fault Injection} & 5.74 & (3.1\%) & 0.20 & (1.9\%)\\
\textbf{Rate Limit} & 8.19 & (4.5\%) & 0.21 & (2.0\%)\\
\textbf{Tap} & 156.09 & (85.0\%) & 2.95 & (8.0\%)\\
\textbf{Lua} & 80.59 & (43.9\%) & 3.18 & (30.2\%)\\
\textbf{WebAssembly} & 26.30 & (14.3\%) & 0.69 & (6.6\%)\\
\bottomrule
\end{tabular}
\vspace{2mm}
\caption{Latency overhead of five filters. The percentage in parentheses denotes the additional overhead atop baseline HTTP mode (without any filters).}
\vspace{-3mm}
\label{tab:filter}
\end{table}

We study five different filters, covering all three ways to write an Envoy filter: 1) Fault Injection: a built-in, C++ filter that helps test the resilience to communication failures; 2) (Local) Rate Limit: a built-in, C++ filter that rate limits traffic to a service instance. 3) Tap (File): a built-in, C++ that records traffic and is configured to log to a file; 4) Lua: a custom, no-op filter written as a Lua script; 5) WebAssemtly: a custom, no-op filter written as a WebAssembly module. We add these filters on Envoy configured in HTTP mode.

\autoref{tab:filter} shows the overhead of each filter inferred by \sys when subjected to the same workload as the previous section (100 byte messages, 30K request per second). We see that different filters have widely different overheads. The baseline overhead of C++ filter is low, as evidenced by the low overhead of Fault Injection and Rate Limit filters. The high overhead of Tap (file) is high because of its interaction with the file system. On the other head, even no-op Lua or WebAssembly filters have substantial latency and CPU overheads, with Lua being 3x more expensive for latency and nearly 5x more expensive for CPU.

To study the composability of filters, we consider five different filter configurations, each with a different way to combine filter types: 1) C$_C$: combines all three types of C++ filters;  2) C$_{LW}$ combines the Lua and WebAssembly filters; 3) C$_{CL}$: combines the Lua filter with all three C++ filters; 4) C$_{CW}$: combines the WebAssembly filter with all three C++ filters; and 5) $C_{CLW}$: combines all five filters. 

\begin{figure}[t]
  \centering
    \subfigure[Latency. \label{fig:filters_latency}]{\includegraphics[width=0.5\linewidth]{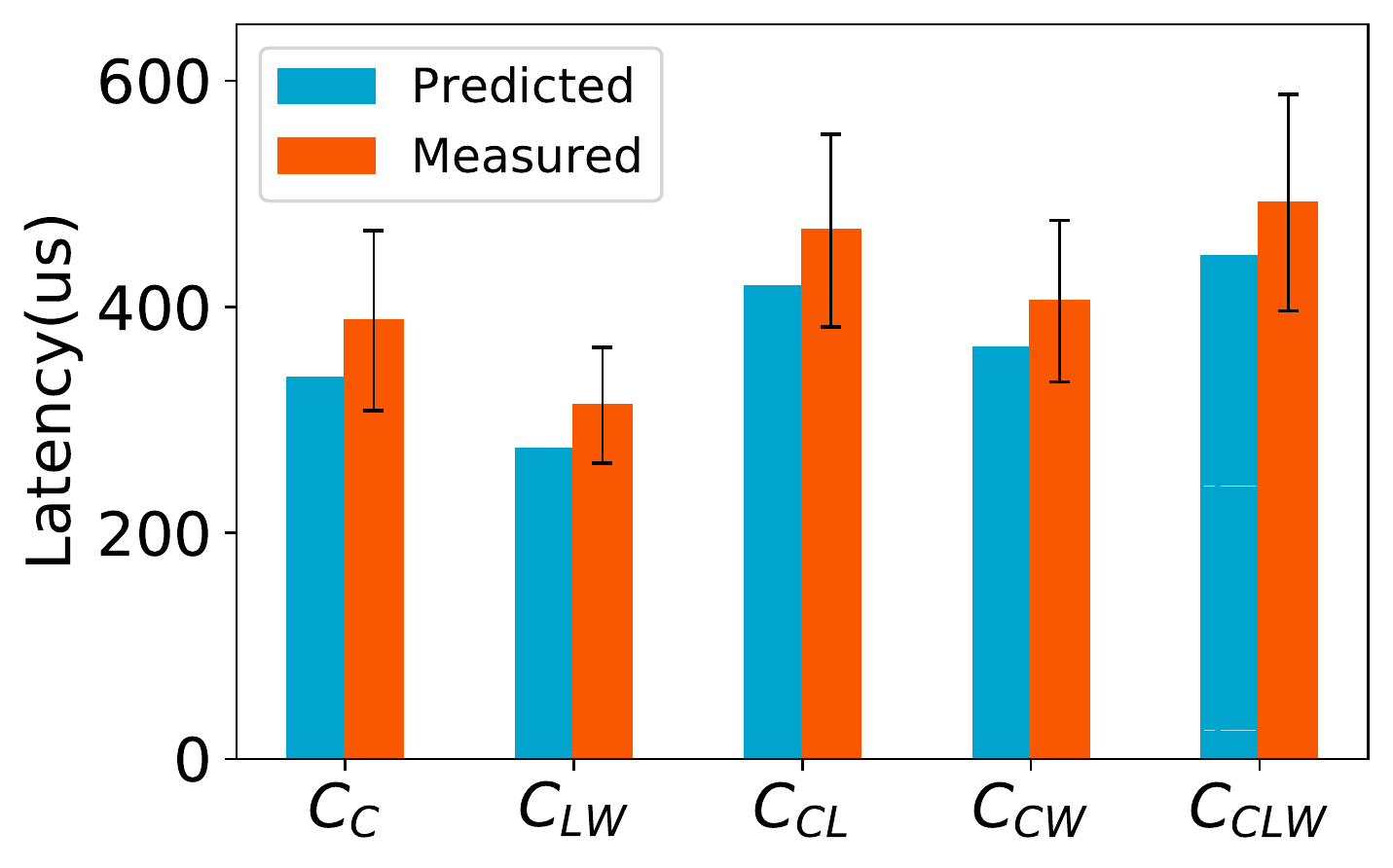}}\hfil 
    \subfigure[CPU Usage. \label{fig:filters_cpu}]{\includegraphics[width=0.5\linewidth]{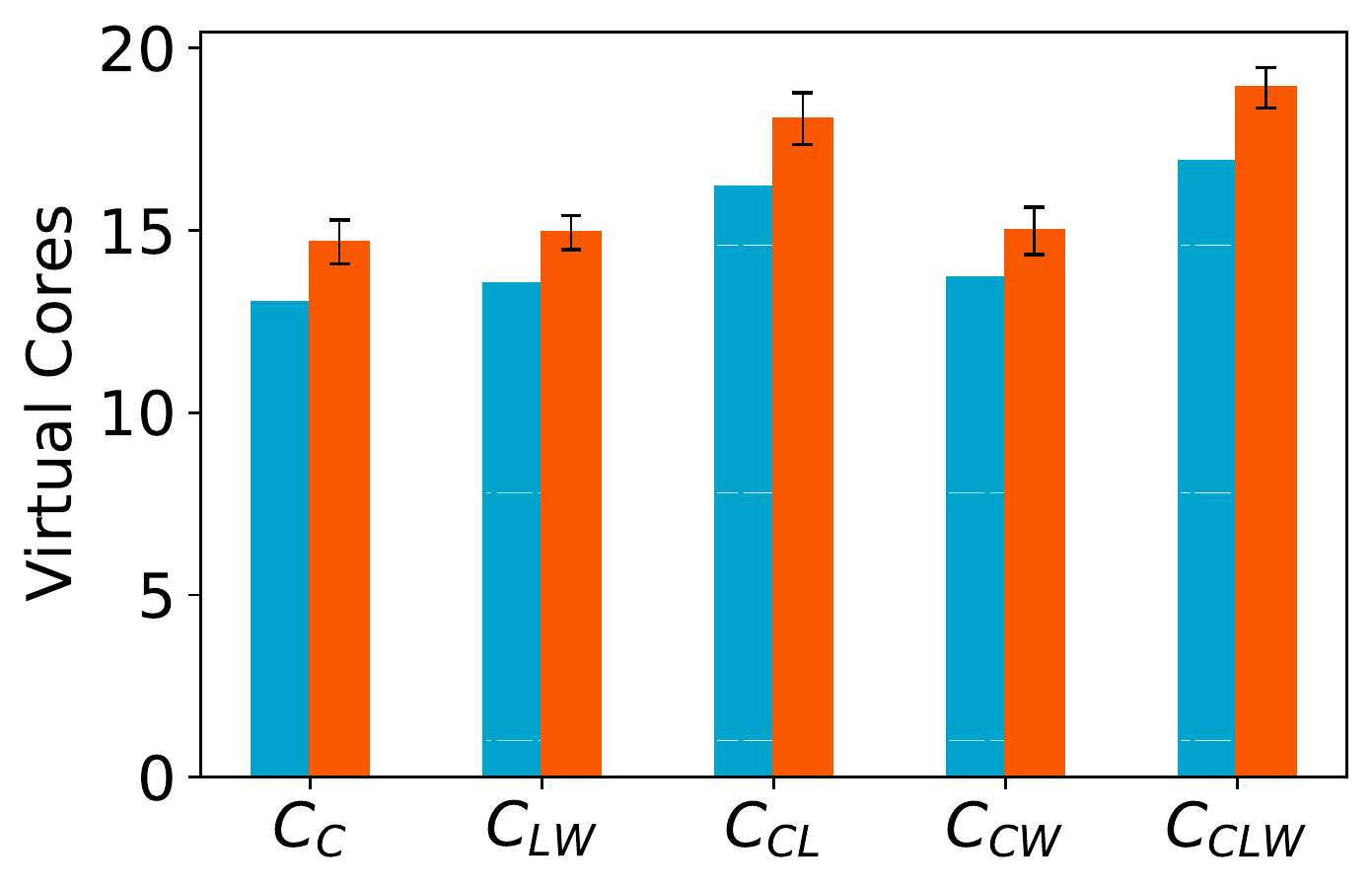}}
  \caption{Prediction results of different filters configurations.}
  \label{fig:filters}
\end{figure}

\autoref{fig:filters} shows both predicted and measured overheads of each of these combinations. The measured overhead denotes latency and CPU usage with the filters minus that without the filters. We see that filter combination overheads can be quite high when multiple expensive filters are employed (something that the developers must avoid). We also see the predictions of \sys, based on adding individual overheads on top of base HTTP proxy overhead in \autoref{tab:filter}, are quite accurate.

\subsection{Impact of Message Size and Rate}
\label{sec:size_and_rate}

We now characterize the impact of message size and rate. We will show that, consistent with our modeling assumptions, the overhead increases with each of these factors. To study the impact of message size, we vary it from 100\,bytes to 16KB. The upper end of this range is well beyond the maximum size that we directly profile (4KB). To study the impact of message rate, we vary it from 10K to 50K requests per second. 

\begin{figure}[t]
\centering
\includegraphics[width=0.55\linewidth]{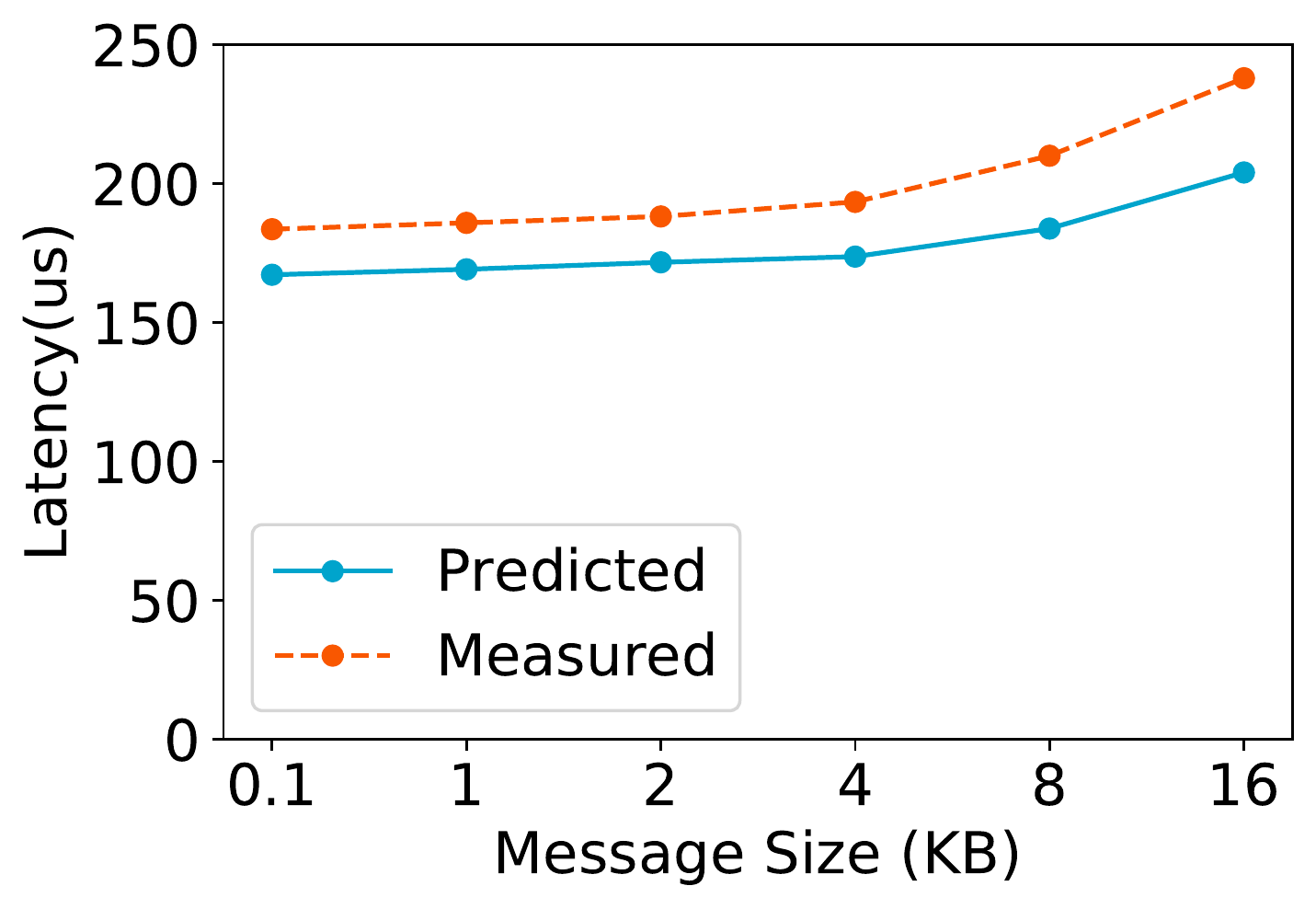}
\vspace{-2mm}
\caption{Impact of message size on service mesh latency overhead. X-axis is on log scale.}
\vspace{-4mm}
\label{fig:latency_model}
\end{figure}

\parab{Latency} \autoref{fig:latency_model} plots latency overhead for HTTP proxy without filters. The latency increase is similar for other protocols. We see that latency overhead increases slowly with message size. Going from 100 bytes to 16~KB (which represents a very large message), the latency overhead increases by 53 \,ms. This increase represents only a 30\% increase for HTTP. The presence of filters does not significantly change the impact of message size on latency, as most filters operate on message headers, not payload (which has the most bytes).

We also see in \autoref{fig:latency_model} that \sys models the impact of latency increase well, though its prediction accuracy drops for very large messages (which is uncommon~\cite{qian2009tcp,montazeri2018homa}). As mentioned earlier (\S\ref{sec:offline}), the reason for this lower accuracy is that messages larger than a few KB are split into multiple units which has higher latency and CPU cost.

\begin{figure}[t]
  \centering
  {
  \subfigure
  {\includegraphics[width=0.48\linewidth,trim=0 0.25cm 0 0]{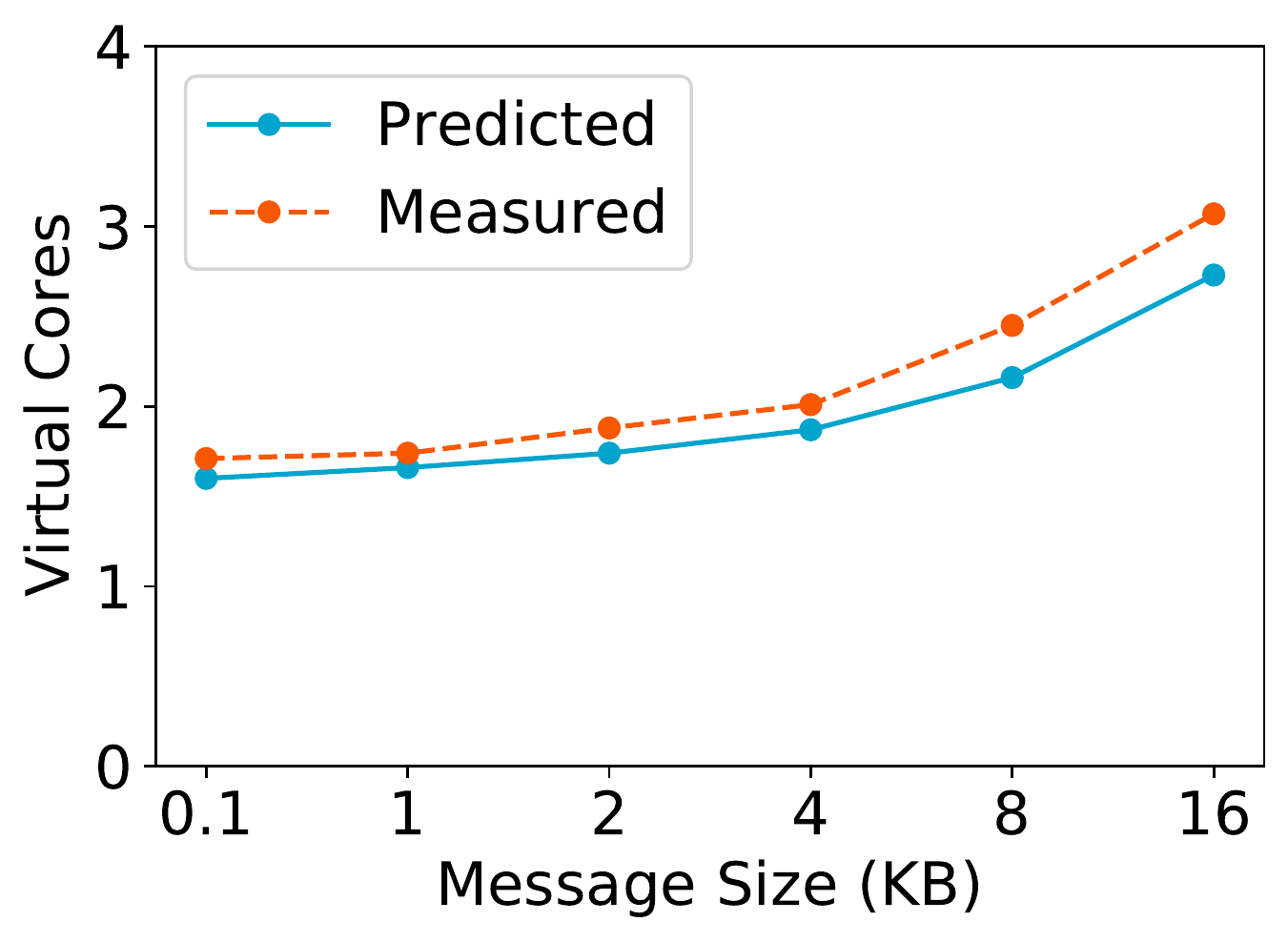}}
  \hfill
  \subfigure
  {\includegraphics[width=0.5\linewidth,trim=0 0.25cm 0 0]{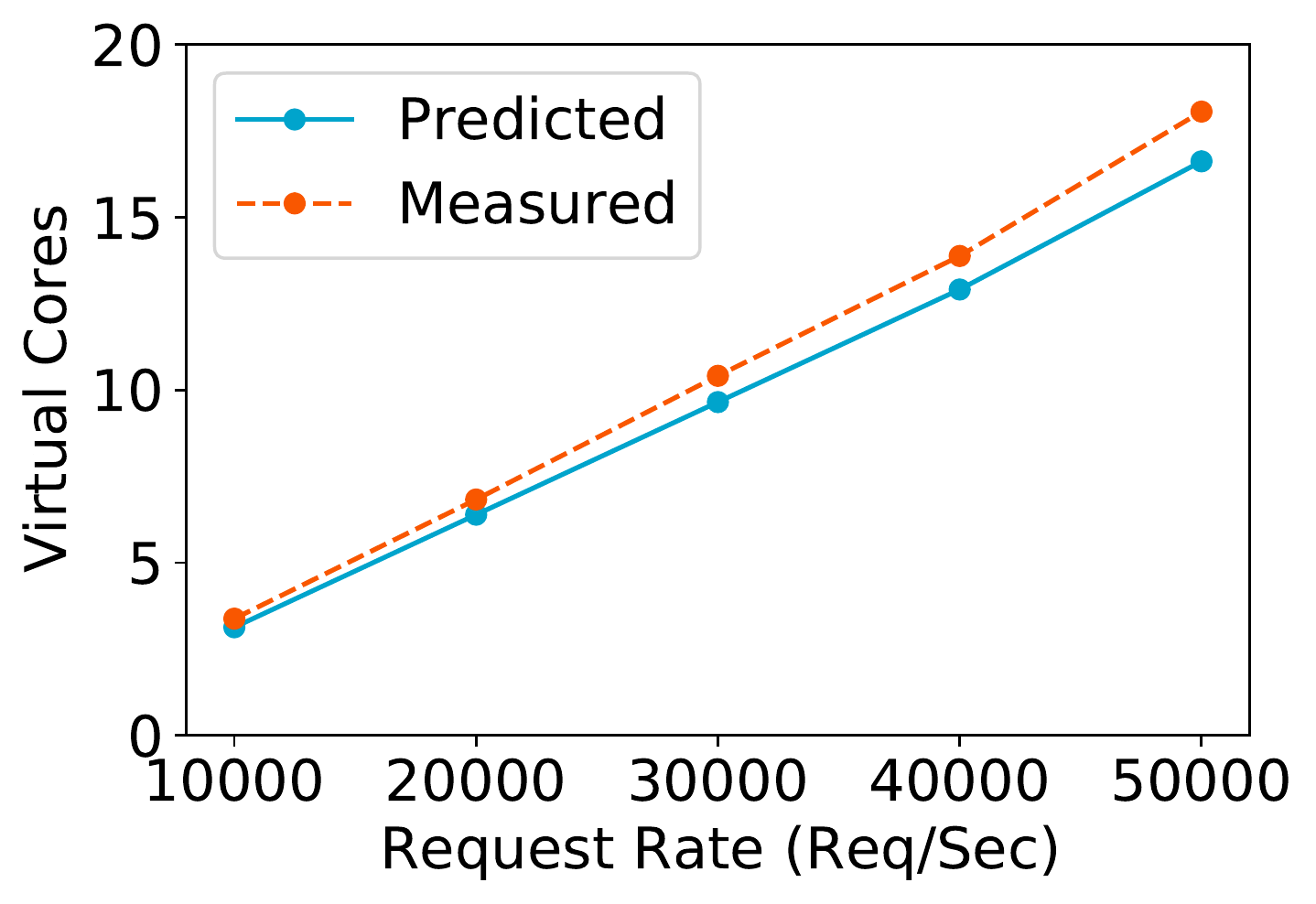}}
  }
  \caption{Impact of message size and rate on service mesh CPU overhead of service meshes.}
  \vspace{-4mm}
  \label{fig:cpu_model}
\end{figure}

\parab{CPU usage.} \autoref{fig:cpu_model} shows CPU usage for HTTP proxy. We see that the CPU overhead increases linearly with message sizes (for a fixed rate) and linearly with message rate (for a fixed size), and that \sys tracks this increase well.

Similar to latency, large messages have relatively low impact on CPU overhead, compared to message rate. CPU usage increases by 44\% when going from 100 bytes to 16 KB. 
\section{Helping Developers Predict Overhead}
\label{sec:casestudy}

In addition to characterizing service mesh overhead in detail, \sys has two more use cases: (1) helping application developers determine how to configure service meshes, and (2) helping service mesh developers evaluate potential optimizations.

\begin{figure}[t]
  \centering
  {
    \subfigure[Absolute latency overhead.\label{fig:latency-diff}]{\includegraphics[width=0.5\linewidth]{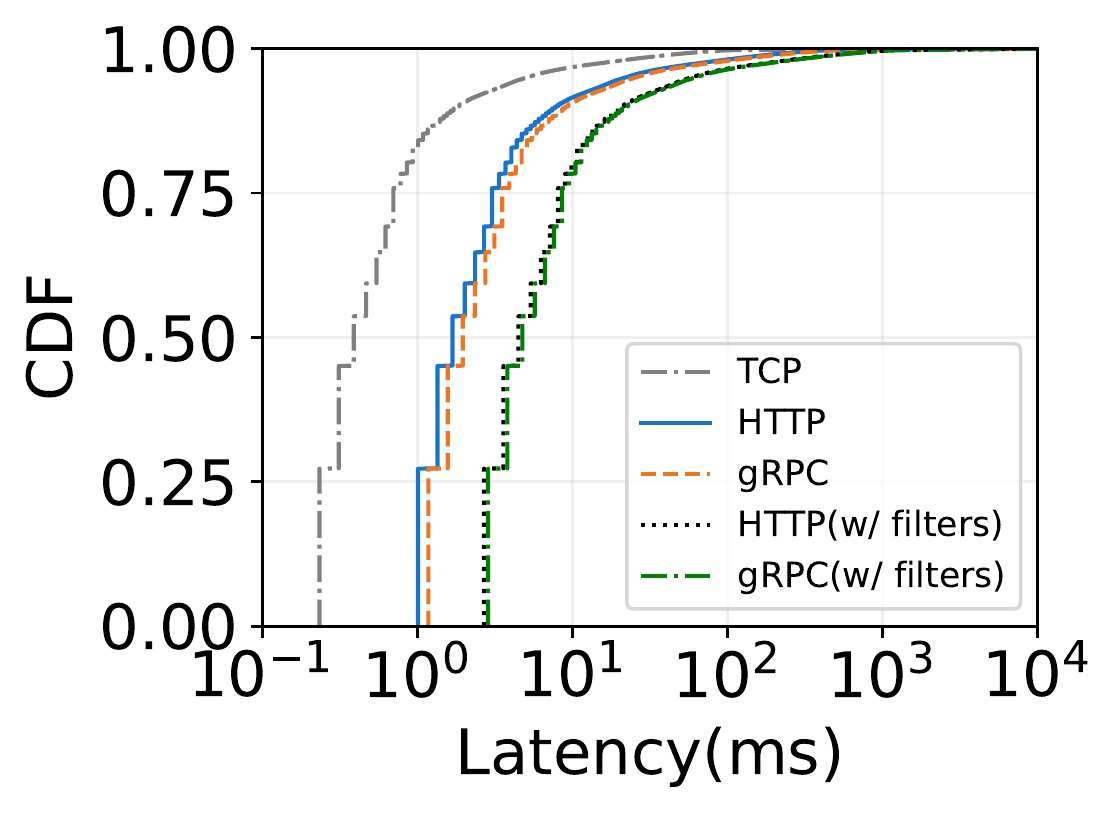}}\hfill
    \subfigure[Absolute CPU overhead. \label{fig:latency-div}]
    {\includegraphics[width=0.5\linewidth]{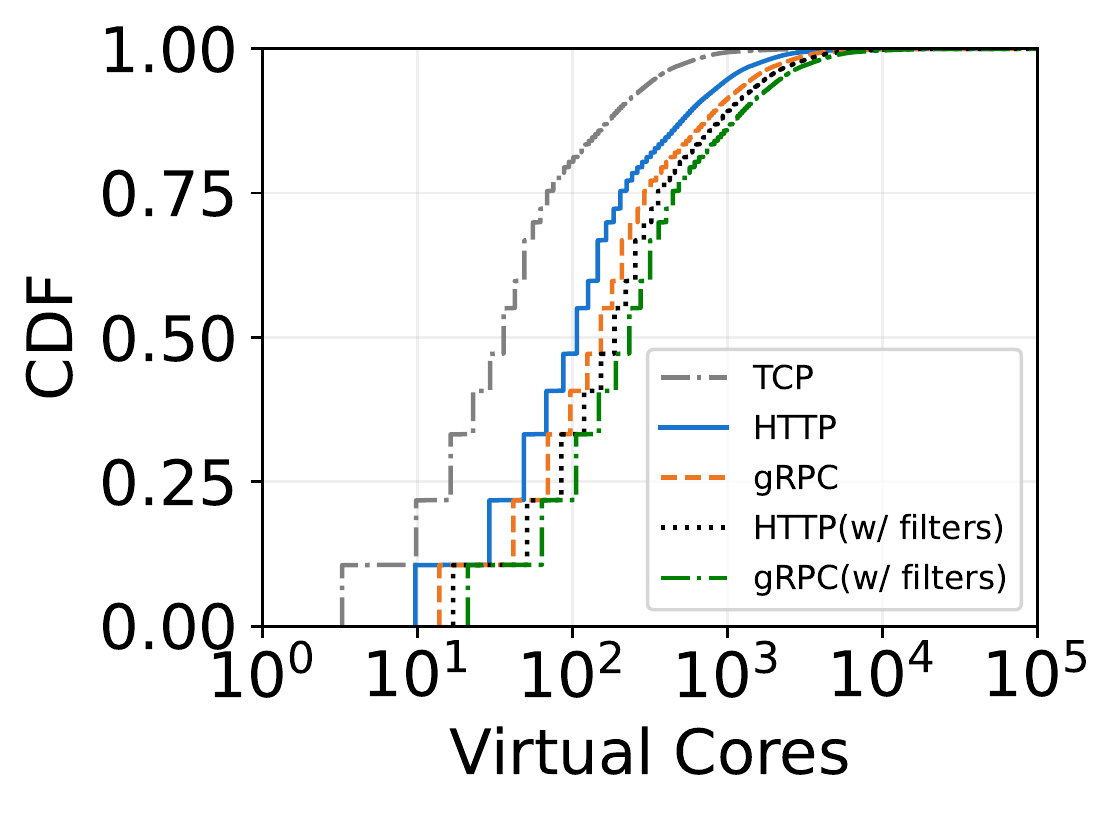}}
  }
  \caption{Latency and CPU overhead for application call graphs in the Alibaba trace.}
  \label{fig:alibaba}
\end{figure}

\subsection{Application Developers}

Given the application call graph, \sys can predict service mesh overhead under different configurations. With this knowledge, they can make the right trade-off between desired functionality and overhead. 

We demonstrate this use case using the Alibaba microservice traces\cite{luo2021characterizing}. These contain over 20M call graphs from microservices-based applications, collected over 7 days in an Alibaba cluster. While most call graphs have a small number of microservices, 10\% of the them have over 40 microservices and the largest ones have thousands of microservices. We randomly select 1M call graphs for our experiments. The traces do not contain message sizes or rates; we assume these to be 100 bytes (because service mesh messages tend to be small) and 30K requests per second which represents moderate load.

We consider five possible service mesh configurations, three protocol modes without filters, HTTP with filters, and gRPC with filters (all five filters described in \autoref{sec:filters}). \autoref{fig:alibaba} shows the latency and CPU overheads. Since a call graph can have multiple paths, the latency overhead is computed for the critical paths, which we extract from the Alibaba trace based on both invocation's timestamp and response time. In any given configuration, the overhead varies by multiple orders of magnitude across applications. Even for simplest configuration (TCP), latency overhead varies from 0.2 to 100 ms and CPU overhead varies from 3 to 1000 virtual cores.  

We also see that the overhead of different service mesh configurations varies by an order of magnitude for both latency and CPU. The median latency overhead is 0.2\.ms in TCP mode but it is 2\,ms in gRPC mode with filters, and the 75th percentile varies from 1 to 10\,ms. Similarly, the median CPU overhead varies from 20 virtual CPU cores in TCP mode to over 200 in gRPC mode with filters.

These massive variations based on service mesh configuration and application characteristics is why we need a tool like \sys using which application developers can learn the overhead of their specific deployment scenarios of interest.

\subsection{Service Mesh Developers}

\sys enables service mesh developers to judge the impact of potential optimizations. There are several ongoing works in the industry today to optimize the service mesh performance overheads~\cite{cillium-sm,istio-ebpf}. While such optimizations can be benchmarked in isolation, it is difficult to understand the end-to-end impact on real-world applications.

To demonstrate this use case, we consider using two Linux kernel features in Envoy. Porting Envoy to use new kernel features is quite a bit of implementation effort (and may also introduce some functional limitations), so the Envoy developers may want to estimate the impact even before taking on this work. For this estimation, all they need to provide \sys is an estimate of new performance profiles for various components. To estimate these speedups we enable these features in the context of a simple sidecar (with 676 lines of C++) that has the same data path architecture as Envoy (\autoref{fig:data_path}). We then profile the components to learn their new performance profiles when these features are active. The Linux features we study are:

\parab{Unix domain sockets}

In \autoref{sec:characterization-breakdown}, we saw that IPC overhead is a significant contributor to overhead, adding at least 11 microseconds to latency and consuming 0.5 virtual cores, representing 30\% of latency overhead and 15\% of CPU overhead in TCP mode. By default, Envoy-to-application IPC uses a TCP connection, which traverses the TCP/IP network stack for the loopback interface. One potential option to reduce overhead is via Unix domain sockets~\cite{uds}, which is lighter weight than TCP sockets while providing the same API.When using a Unix domain socket, the kernel copies the data to kernel space and directly puts the constructed socket buffer on the receiving side's socket queue, avoiding the expensive network stack processing. 

\begin{figure}[t]
  \centering
  {
    \subfigure[Unix Domain Socket. \label{fig:latency-div-uds-100B}]{\includegraphics[width=0.5\linewidth]{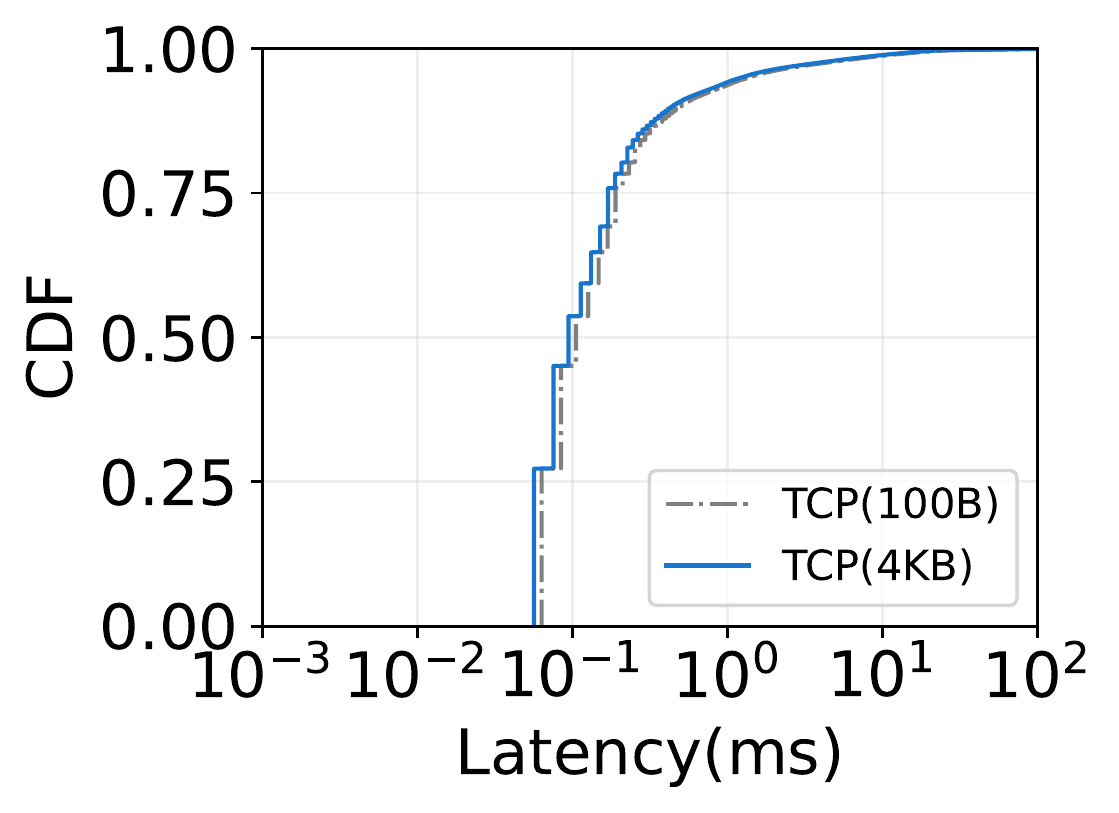}}\hfill
    \subfigure[Zero Copy Writes. \label{fig:latency-div-uds-4KB}]{\includegraphics[width=0.5\linewidth]{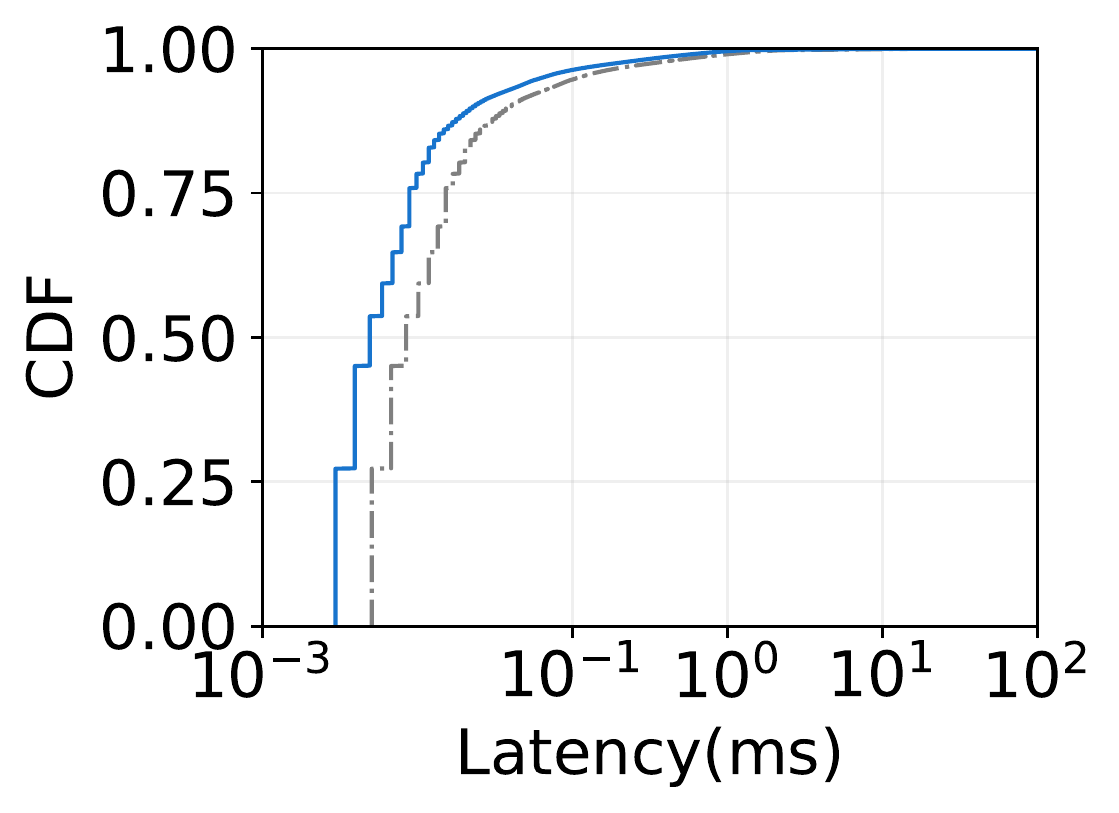}}
  }
  \caption{End-to-end latency reduction in the Alibaba trace with Linux features.}
  \label{fig:alibaba-latency-opt}
\end{figure}

\begin{figure}[t]
  \centering
  {
    \subfigure[Unix Domain Socket. \label{fig:latency-div-zero-copy-100B}]{\includegraphics[width=0.5\linewidth]{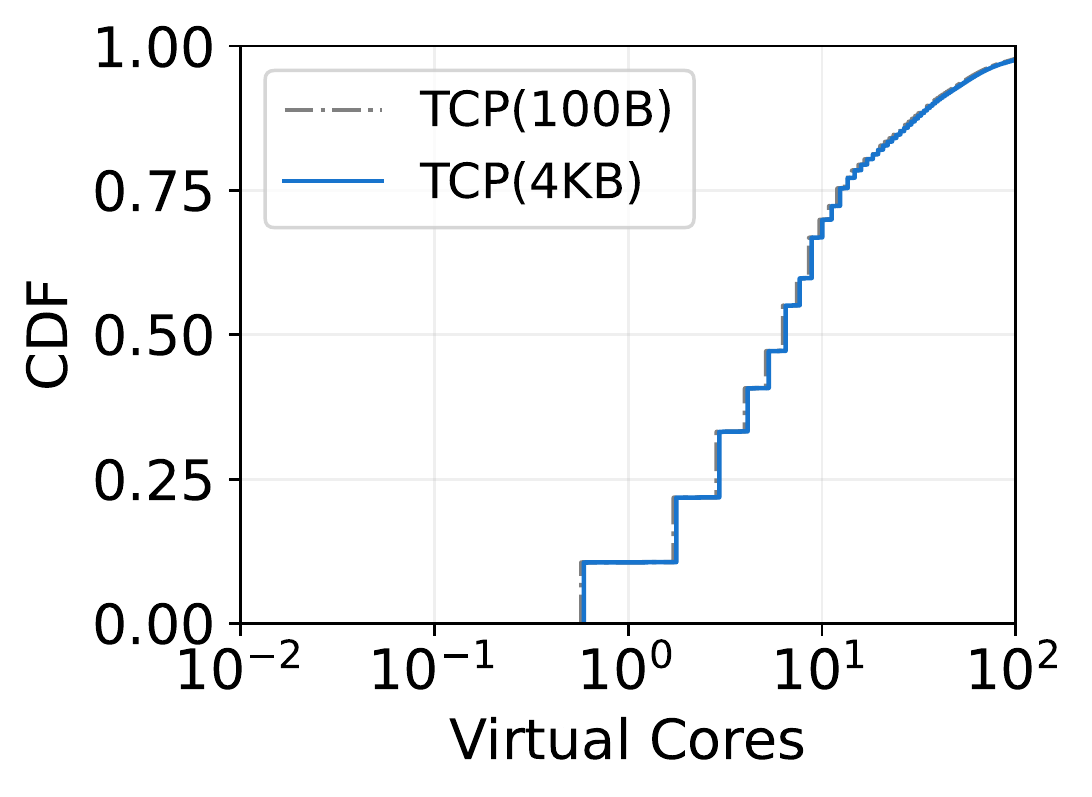}}\hfill
    \subfigure[Zero Copy Writes. \label{fig:latency-div-zero-copy-4KB}]{\includegraphics[width=0.5\linewidth]{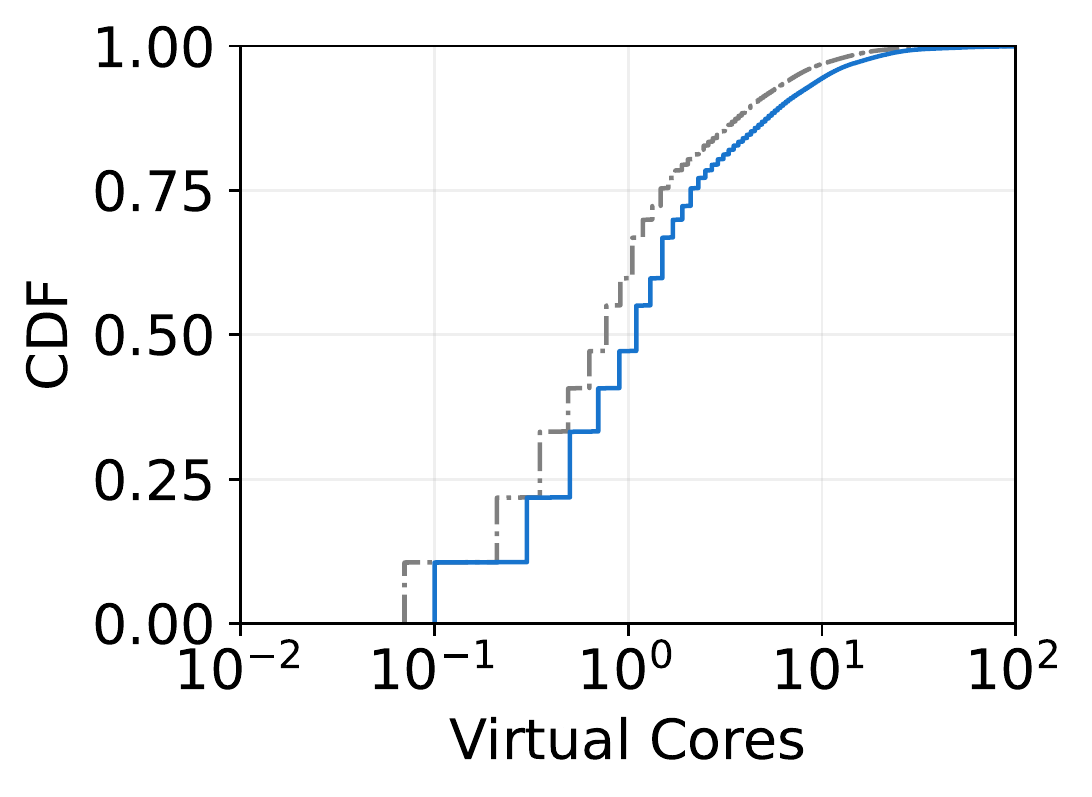}}
  }
  \caption{End-to-end CPU usage reduction in the Alibaba trace with Linux features.}
  \label{fig:alibaba-cpu-opt}
\end{figure}

\parab{Reducing Write Overhead}
A second significant source of Envoy overhead is the latency and CPU usage of copying data, which is embedded in the Read and Write component. It is already noticeable with 100 byte messages (\autoref{tab:breakdown-100B}) and gets worse with larger messages. Linux kernel supports zero-copy TCP sockets starting from version 4.14. For write calls, applications can pin memory buffers in userspace; the kernel signals to the application after sending the buffers to enable garbage collection or re-use of the buffers. This eliminates the need for copying the data from userspace to the kernel. Linux does not support zero-copy for read calls.

We use \sys to evaluate the performance implications of these optimizations using the same Alibaba microservice workloads as above. We consider TCP mode sidecars. The absolute savings for HTTP or gRPC mode will be similar but the relative advantage in those modes will be small because their performance is bottlenecked by parsing rather than IPC or data copy. 

\autoref{fig:alibaba-latency-opt} shows the latency overheads when using Unix domain socket and zero copy write. We observe substantial improvement from Unix domain socket when message size is 100\,Bytes: the average speedup is 0.71\,ms. For 4KB messages, the average speedup 0.78\,ms. However, because the latency prior to the optimization is higher (4.01\,ms versus 2.88\,ms), the relative decrease in latency is lower with 4KB. This result is in line with IPC overheads constituting less to the total latency overheads when message size is large. 

We also see that zero copy writes bring negligible performance improvement in our setting. This optimization has additional performance cost for buffer lifecycle management~\cite{msgzerocopy}. For message size regime of most interest for microservices, the gains of avoiding a copy is negated by this additional cost. 

\autoref{fig:alibaba-cpu-opt} shows the CPU overheads with the two optimizations. When messages are 100\,bytes, the average CPU overheads reduces from 86 to 71 virtual cores with Unix domain sockets. The difference is quite substantial. When messages are 4KB, Unix domain sockets reduce the CPU overheads from 107 to 91 virtual cores. 
As for latency, zero copy writes do not improve the CPU overheads notably, reducing the CPU overheads from 86\,ms to 84\,ms for 100-byte message . 

\section{Discussion and Future Work}
\label{sec:discussion}

Our work provides a systematic way to understand the overhead of service meshes and confirms that they can significantly increase latency and CPU usage of distributed, microservices applications. There are several ongoing works in the industry targeting at reducing service mesh performance overheads, and our work brings new insights on them.

\parab{Will In-Kernel Sidecars Reduce Service Mesh Overheads?} The Linux kernel has increasing support for extensibility using eBPF. Researchers have also proposed using safe languages, such as Rust, for kernel extension development~\cite{kernel-rust, miller2021bento}. One approach to reduce the system call overheads and the data copy overheads (across userspace and kernel) is to implement a sidecar's functionality inside the kernel. Katran~\cite{katran} offloads layer-4 load balancers into the kernel using BPF. It is now possible for Envoy to run a limited set of filters directly in Cillium~\cite{cillium}, a popular framework for using eBPF on the network data path.  However, our study shows that while removing the system call and data copy overhead can be useful for TCP proxies, it will offer limited performance improvement for HTTP and gRPC proxies because protocol parsing is the major overheads in those configurations. 

\parab{Will Hardware Offloading Reduce Service Mesh Overheads?} Another direction researchers are currently exploring is to offload the sidecar logic to a programmable network hardware~\cite{cui2021smartnic}. While this is a promising direction, there are substantial challenges. For example, it is still questionable if programmable network hardware can do complex layer-7 protocol processing efficiently. There are a variety of layer-7 protocols (e.g., HTTP, gRPC, MySQL). These protocols are more complex than the fixed functions people typically offload to programmable network hardware, such as firewalls, NAT, and layer-4 load balancers~\cite{miao2017silkroad, zhang2020gallium, li2016clicknp}. In addition, many filters in sidecars (e.g., encryption) require reconstructing the original data stream, and this means programmable network hardware also needs to run TCP/IP packet processing (e.g., packet loss recovery, congestion control). Prior works on offloading application logics to programmable network hardware use UDP as the transport to circumvent this issue~\cite{mangpo2018floem}.

\parab{New Directions on Reducing Service Mesh Performance Overheads.} Our work shed the light on some new directions that worth exploration. We observe that protocol parsing is a major overhead for HTTP and gRPC proxies. Unfortunately, using a sidecar today means that there is duplicated protocol processing. A sidecar parses an HTTP request from TCP streams, optionally modifies it, and serializes it into an TCP stream again.
The application service receiving this stream has to parse the HTTP request yet again. 

There are two potential methods to eliminate this double parsing. The first is by linking sidecars to libraries that applications use to parse various protocols. Another is to create a new transport protocol for efficient parsing by sidecars, which is possible in this context where both ends of the communication are sidecars. Both of these methods, however, have limitations. The first one does not work with unmodified applications and the second one does not help when filters need to inspect HTTP headers added by applications. In future work, we will investigate these and other methods.

\section{Related Work}

Our work is related to several threads of prior work.

\parab{Performance of the host network stack.} 
Understanding the host networking stack performance is a common goal in many previous works. Peter et al.~\cite{peter2014arrakis} breaks down the latency overheads of Linux network stack.
Neugebauer et al.~\cite{neugebauer2018pcie} studies how PCIe affects network performance in host networking. Farshin et al.~\cite{farshin2020ddio} examines how Intel Data Direct I/O technology (for NIC to access CPU's last-level cache directly) speeds up host networking performance. More recently, Nsight~\cite{haecki2022nsight} uses Intel Processor Tracing to diagnose latency in network stacks. 

While we share data gathering primitives from these works, our focus is on the data path of service meshes (which traverses the network stack multiple times and has a substantial userspace processing component). We decompose the overhead of a sidecar proxy in the datapath of host networking, and we identify the key contributors to high overhead such as IPC and protocol parsing.

\parab{Performance of network proxies.} 
Many works have investigated the performance of network proxies and developed improvements such as hardware offloading~\cite{zhang2020gallium, miao2017silkroad, pontearelli2019flowblaze} and re-homing TCP connections~\cite{hayakawa2021prism}. However, these works mostly focus on layer-4 network proxies, while sidecars are layer-7 proxies. Our measurements of sidecars provide insights into performance bottlenecks of layer-7 proxies. In the future, we plan to combine insights from our study and techniques for improving the performance of layer-4 proxies to develop high-performance layer-7 proxies.

Maltzahn et al.~\cite{maltzahn1997performance} study the performance characteristics of Web proxies. The performance of Web proxies mostly depends on how proxies cache web contents, where sidecars' performance depends on IPC and buffer parsing.

\parab{Reducing inter-process communication overheads.} Reducing IPC overheads is one of the oldest research topics in the operating system community. Immich et al.~\cite{immich2003performance} and Venkataraman et al.~\cite{venkataraman2015evaluation} study the existing IPC mechanisms' performance on Linux. IPC performance is a critical design aspect for microkernels~\cite{liedtke1993improving, elphinstone2013l4}. The goal of our work is not to develop techniques to lower IPC overhead but to build a tool that helps evaluate the impact of such techniques on the end-to-end performance of service meshes.

\section{Conclusion}

\sys is a tool to systematically quantify the overhead of service meshes. Its decompositional approach can analyze a wide range of deployment scenarios (i.e., the combination of service mesh configuration and application characteristics), without the need to directly measure them (which would be intractable). This ability can help application developers pick the appropriate service mesh configuration for their specific application needs; as we showed using a large dataset of microservice applications, the overhead of service meshes can vary by orders of magnitude based on the configuration, and in a given configuration, the overhead can again vary by orders of magnitude across applications. 

\sys can also identify the primary contributors to the overhead in any scenario. We find, for instance, that IPC and socket writes are the main contributors when the service mesh is configured in TCP mode but protocol parsing dominates in other modes. Our tool and findings can thus also help service mesh developers as they work to lower the overhead of service meshes, which are now a central component of the modern application ecosystem.

\bibliographystyle{plain}%
\bibliography{ref}

\end{document}